\documentclass[%
 reprint,
superscriptaddress,
nofootinbib,
 amsmath,amssymb,
 aps,
 prd,
]{revtex4-1}

\usepackage{graphicx}
\usepackage{dcolumn}
\usepackage{bm}

\usepackage[colorlinks]{hyperref}
\usepackage{tabularx}
\usepackage{graphicx}
\usepackage{amssymb,bm,tensor}
\usepackage{xcolor}
\usepackage{amsmath}
\usepackage[varg]{txfonts}
\usepackage{enumerate}
\usepackage[normalem]{ulem}
\usepackage{bm}
\usepackage[OT1]{fontenc}
\usepackage{physics}

\newcommand{\mnras}{Mon. Not. R. Astron. Soc.}
\newcommand{\sect}{Sec.~}
\newcommand{\fig}{Fig.~}

\newcommand{\eq}{Eq.~}
\newcommand{\eqs}{Eqs.~}
\newcommand{\tab}{Table~}
\newcommand{\tabs}{Tables~}
\newcommand{\pystgromx}{\texttt{pySTGROMX}}

\newcommand{\MG}[2]{#2}
\newcommand{\MMG}[2]{#2}

\begin{document}

\preprint{PRD}

\title{Extended reduced-order surrogate models for scalar-tensor gravity in the
strong field and applications to binary pulsars and gravitational waves}

\author{Minghao Guo}
\email[Corresponding author: ]{gmh@pku.edu.cn}
\affiliation{Peking University, Beijing 100871, China}
\affiliation{Department of Astrophysical Sciences, Princeton University, Princeton, NJ 08544, USA}
\author{Junjie Zhao}%
\affiliation{School of Physics and State Key Laboratory of Nuclear Physics and Technology, Peking University, Beijing 100871, China}
\author{Lijing Shao}%
\email[Corresponding author: ]{lshao@pku.edu.cn}
\affiliation{Kavli Institute for Astronomy and Astrophysics, Peking University, Beijing 100871, China}
\affiliation{National Astronomical Observatories, Chinese Academy of Sciences, Beijing 100012, China}

\date{\today}

\begin{abstract}
\MG{We investigate the scalar-tensor gravity of Damour and Esposito-Far\`ese
(DEF) with spontaneous scalarization phenomena developed for neutron stars
(NSs).  Instead of solving the modified Tolman-Oppenheimer-Volkoff equations for
slowly rotating NSs via the shooting method, we construct a reduced-order
surrogate model to predict the relations of mass, radius, moment of inertia,
effective scalar coupling, and two extra coupling parameters of a NS to its
central matter density. We code the model in the \pystgromx\ package that speeds
up the calculations at two and even three orders of magnitude and yet still
keeps accuracy at $\sim1\%$ level. Using the package, we can predict
\textit{all} the post-Keplerian parameters in the timing of binary pulsars
conveniently, which allows us to place comprehensive constraints on the DEF
theory in a quick and sound way. As an application, we perform Markov-chain
Monte Carlo simulations to constrain the parameters of the DEF theory with
well-timed binary pulsars.\\}{Statistically sound tests of scalar-tensor gravity
theories in the strong-field regime usually involves computationally intensive
calculations.  In this study, we construct a reduced order surrogate model for
the scalar-tensor gravity of Damour and Esposito-Far\`ese (DEF) with spontaneous
scalarization phenomena developed for neutron stars (NSs). This model allows us
to perform a rapid and comprehensive prediction of NS properties, including
mass, radius, moment of inertia, effective scalar coupling, and two extra
coupling parameters. We code the model in the \pystgromx\ package, as an
extension of our previous work, that speeds up the calculations at two and even
three orders of magnitude and yet still keeps accuracy of $\sim1\%$. Using the
model, we can calculate \textit{all} the post-Keplerian parameters in the timing
of binary pulsars conveniently, which provides a quick approach for us to place
comprehensive constraints on the DEF theory. We perform Markov-chain Monte Carlo
simulations with the model to constrain the parameters of the DEF theory with
well-timed binary pulsars.} Utilizing five NS-white dwarf and three NS-NS
binaries, we obtain the most stringent constraints on the DEF theory up to now.
Our work provides a public tool for quick evaluation of NSs' derived parameters
to test gravity in the strong-field regime.
\end{abstract}

\maketitle

\section{\label{sec:intro}Introduction}

Albert Einstein's theory of general relativity (GR)~\cite{Einstein:1915ca}
remains the most accurate theory of gravity for more than a century.  This
elegant theory has passed all tests with flying colors from, e.g., the Solar
System experiments~\cite{Will2014LRR....17....4W}, cosmological
observation~\cite{Clifton:2011jh}, the timing of binary
pulsars~\cite{Stairs2003LRR.....6....5S, Wex:2014nva, Shao:2016ezh}, and
gravitational waves (GWs) from coalescing binary black holes
(BBHs)~\cite{TheLIGOScientific:2016src, Abbott:2017vtc,
LIGOScientific:2019fpa,LIGOScientific:2020tif} and binary neutron stars
(BNSs)~\cite{TheLIGOScientific:2017qsa, GBM:2017lvd,Abbott:2018lct}.  From the
Earth up to the Universe, from weak to strong gravitational field, GR remains
the gold standard.

There are good theoretical reasons to go beyond GR,
however~\cite{Berti:2015itd}.  Therefore, even with the success of GR,
considerable efforts are still being made for alternative theories of gravity
(see Refs.~\cite{Will2014LRR....17....4W,Will:2018bme} for a review). In GR,
gravity is mediated solely by a massless, spin-2 tensor field, namely the metric
of spacetime $g_{\mu\nu}$. Differently from GR, scalar-tensor theories of
gravity, as natural \MG{}{and well-motivated} alternatives, add one or more extra scalar degrees of
freedom in the gravitational sector. They not only arise naturally as a possible
low-energy limit of higher dimensional theories, such as Kaluza-Klein
theory~\cite{Kaluza:1921tu,Klein1926ZPhy...37..895K} and string
theories~\cite{Fujii:2003pa}, but also have a potential connection to the
inflation, the dark energy, and a yet unknown unified theory of quantum
gravity~\cite{Clifton:2011jh}.  Originally suggested by Scherrer in
1941~\cite{Goenner:2012cq}, the most popular scalar-tensor gravity theories are
developed in a modern framework by Jordan~\cite{Jordan:1949zz,Jordan:1959eg},
Fierz~\cite{Fierz:1956zz}, Brans and Dicke~\cite{Brans:1961sx} (JFBD; see a
review in Ref.~\cite{Fujii:2003pa}).  JFBD-like theories, as metric theories of
gravity, do not violate the weak equivalence principle but the strong
equivalence principle (SEP) due to the nonminimally coupled scalar field in the
Einstein-Hilbert action~\cite{Shao:2016ezh, Will:2018bme}. Tests of SEP, which
is the heart of GR, provide a powerful tool to experimentally constrain these
theories.

In this paper, we restrict our attention to \MG{a class of special JFBD-like
gravity formulated}{a category of JFBD-like theory proposed}
 by Damour and Esposito-Far\`ese
(DEF)~\cite{Damour1992CQGra...9.2093D, Damour1993PhysRevLett.70.2220,
Damour1996PhysRevD.54.1474}, where prominent violations of the SEP due to
\textit{nonperturbative strong-field effects} are known to arise.  The DEF
theory can pass present weak-field gravitational tests, such as the Cassini
experiment~\cite{Bertotti:2003rm}, but \MG{}{still} exhibits very significant strong-field
deviations away from GR in the systems involving strongly self-gravitating
neutron stars (NSs)~\cite{Damour1993PhysRevLett.70.2220,
Shao2017PhRvX...7d1025S}.

\MG{In the weak-field regime, scalar-tensor theories have been extensively
investigated, mostly from experiments in the Solar System, like the Cassini
probe and other space missions~\cite{Bertotti:2003rm}. In the parametrized
post-Newtonian (PPN) framework, it is confirmed to a high precision
$\sim10^{-5}$ for the Eddington-Robertson-Schiff parameters~\cite{Will:2018bme},
indicating that the DEF theory is compatible with the weak-field tests when its
weak-field coupling parameter, $\alpha_0$, is small.\\}{In the weak-field regime, observations in the Solar System, such as the Cassini
probe~\cite{Bertotti:2003rm}, have placed stringent bounds on scalar-tensor
theories. Shapiro time-delay measurements with the Cassini spacecraft have confirmed to a high precision
$\sim10^{-5}$ that the Eddington-Robertson-Schiff parameters are in agreement with GR prediction in the parametrized
post-Newtonian (PPN) framework~\cite{Will:2018bme}. Nevertheless, the DEF theory is compatible with the weak-field tests as long as its
weak-field coupling parameter, $\alpha_0$, is small.} 

However, in the strong-field regime for the DEF theory, \MG{}{even with a small $\alpha_0$,} one kind of
nonperturbative strong-field effects, the so-called \textit{spontaneous
scalarization}, occurs as a scalar analogue of the phase transition in
ferromagnetism~\cite{Damour1993PhysRevLett.70.2220, Damour1996PhysRevD.54.1474,
Sennett:2017lcx}.  It naturally arises in an isolated compact star such as a NS
under a certain condition with the scalar field excited far above its background
value.  Such deviation from GR introduces considerable modifications into the
properties of scalarized compact stars. It also affects the relativistic orbital
motion if a scalarized NS is in a binary, via, e.g., a body-dependent effective
gravitational constant, extra gravitational binding energy related to the scalar
field, and dipolar radiation in addition to the canonical quadrupolar radiation
in GR~\cite{Damour1992CQGra...9.2093D, Damour2009ASSL..359....1D}. Long-term
monitoring of binary pulsars and transient observations of GWs from BNS
coalescences are therefore powerful tools to probe the contribution from
spontaneous scalarization in extreme environments of strong gravitational
fields~\cite{Wex:2014nva, Damour1996PhysRevD.54.1474, Anderson:2019eay}.

The high-precision timing of binary pulsars provides some of the tightest
gravity tests with strongly self-gravitating bodies in the quasi-stationary
strong-field gravity regime~\cite{Stairs2003LRR.....6....5S, Wex:2014nva,
Shao:2016ezh, Shao:2019nso}. In this regime, gravitational fields are strong
with large spacetime curvature in the vicinity of the NSs while the typical
velocity $v$ is much smaller than the speed of light $c$ with a ratio of
$v/c\sim 10^{-3}$.  To extract the information from pulsar timing for testing
gravity theories, the parametrized post-Keplerian (PPK) formalism was \MG{developed
as a generic pulsar timing model}{constructed as a general framework}~\cite{Damour1992PhysRevD.45.1840}. The
dynamical information can be obtained from the pulsar timing and pulse-structure
data by fitting the data to a model consisting of \MG{a set of}{the} theory-independent
Keplerian and post-Keplerian \MG{timing}{} parameters. \MG{The PPK formalism is a powerful
tool and makes it possible to provide constraints on a variety of alternative
gravity theories.}{}

In the DEF theory, one of the PPK parameters,  $\dot{P}_b$, which is related to
binary orbital decay, is modified due to the extra dipolar
radiation~\cite{Damour1992CQGra...9.2093D}. The dipolar contribution,
corresponding to a $-1$ post-Newtonian (PN) correction,\footnote{We refer a
correction at $n$\,PN order to $O(v^{2n}/c^{2n})$ modification relative to the
Newtonian order. For GWs, the quadrupolar radiation is denoted as $0$\,PN, and
the dipolar radiation is at $-1$\,PN.} may dominate the radiation when $v/c$ is
small.  Thus, at the early time of the binary system, due to a small
$v/c\sim10^{-3}$, this mechanism enhances the energy flux of GWs emitted from
the system and thus changes $\dot{P}_b$ in a noticeable way.  In addition, the
DEF theory predicts that, due to the spontaneous scalarization, \textit{all} the
PPK parameters are modified from the GR
prediction~\cite{Damour1992PhysRevD.45.1840}\MG{, thus}{. Therefore,} these parameters can be
combined to constrain the DEF theory~\cite{Damour1996PhysRevD.54.1474}.

With the detection of GW events, especially coalescing BNSs,
GW170817~\cite{TheLIGOScientific:2017qsa} and a possible candidate
GW190425~\cite{LIGOScientific:2020ibl} up to date, we have a new testbed in
probing the strong-field gravity\MG{}{ in highly dynamical regime}. \MG{In the late inspiral stage, the spacetime of
BNSs is strongly curved and highly dynamical in the vicinity of NSs, providing a
powerful laboratory in the highly dynamical strong-field regime.}{} Matched-filter
analyses, used in parameter estimation of GWs~\cite{Finn:1992wt}, are sensitive
to GW phase evolution, which is modified in the DEF theory and can be
distinguished from GR. Thus, GW signals can also be utilized to bound the DEF
theory.

So far, however, \MG{limited by the sensitivity of the LIGO/Virgo detectors below
tens of hertz,}{} the bound on dipolar radiation from \MG{GW170817}{GW of coalescing BNSs} is still looser than
that from the timing of binary
pulsars~\cite{Shao2017PhRvX...7d1025S,Zhao2019PhysRevD.100.064034}, since
dipolar radiation corresponds to a $-1$ PN correction and plays a relatively
important role when $v/c\ll1$\MG{}{, corresponding to low-frequency signals, where sensitivity of the LIGO/Virgo detectors is limited}. Future ground-based and space-based GW detectors
with a better low-frequency sensitivity, such as Cosmic Explorer
(CE)~\cite{Evans:2016mbw-CE}, Einstein Telescope (ET)~\cite{Hild:2010id-ET},
DECi-hertz Interferometer Gravitational wave Observatory
(DECIGO)~\cite{Yagi:2009zz} and Decihertz Observatory
(DO)~\cite{Sedda2020CQGra..37u5011A, Sedda:2021hpg} will place tighter
constraints on gravity theories by either \MG{extending the sensitivity bands to be
below $10\,{\rm Hz}$}{enabling observations below $10\,{\rm Hz}$} or increasing the sensitivity further (see e.g.
Ref.~\cite{Liu:2020nwz}). 

To perform tests on the scalar-tensor gravity, first, one must derive the
relevant predictions on observations precisely.  The predictions on some
properties of a  NS, such as its radius $R$, mass $m_A$, moment of inertia
$I_A$, and the scalar coupling parameter $\alpha_A$, are derived by numerically
integrating the modified Tolman-Oppenheimer-Volkoff (mTOV) equations of a slowly
rotating NS with the shooting method~\cite{Damour1993PhysRevLett.70.2220,
Damour1996PhysRevD.54.1474}. This integration depends on the equation of state
(EOS) of NS matters, which is\MG{}{, unfortunately,} still full of large uncertainties (see e.g.
Refs.~\cite{Lattimer:2000nx, Shao:2019gjj}). 

Apart from the NSs' properties above, \MG{many PPK parameters in binary pulsars,
such as the periastron advance rate and Einstein delay parameter, require other
derived properties of NSs in the DEF theory}{some other
properties in the DEF theory are also required to predict some of PPK parameters in binary pulsars}. They consist of the coupling
parameters $\beta_A$ and $k_A$, which are \MG{predicted}{derived} by calculating the
derivatives of scalar coupling $\alpha_A$ and moment of inertia $I_A$, with
respect to the scalar field at infinity~\cite{Damour1996PhysRevD.54.1474}. Note
that the calculations of $\beta_A$ and $k_A$ should be performed for a fixed
value of baryonic mass $\bar{m}_A$. This procedure requires \MG{}{one to perform} the shooting method
for both $\bar{m}_A$ and the scalar field
simultaneously~\cite{Damour1996PhysRevD.54.1474}.  The calculations for such
parameters are very time-consuming and thus expensive for large-scale
computation.

In practice, to constrain the free parameters in the DEF theory \MG{with the
well-timed pulsars}{} in a statistically sound way, we use Bayesian inference
through Markov-chain Monte Carlo (MCMC) simulations. This approach \MG{requires}{involves} the
evaluation of the likelihood function hundreds of thousands to millions of
times\MG{.  Each step of MCMC simulation requires the corresponding NS properties
derived from the integration of the mTOV equations with the shooting method.}{ with solving the mTOV equations each time.} The
whole simulation is thus time-consuming and expensive. Such computationally
intensive studies have been conducted in Ref.~\cite{Shao2017PhRvX...7d1025S} for
the first time.

In this study, we avoid solving the mTOV equations iteratively and repeatedly
by trial and error \MMG{}{during the data analysis process.} \MMG{, but to}{Instead, we} build a reduced order surrogate model (ROM) \MMG{to predict NS
properties}{in advance with the existing mTOV solutions}.
\MMG{}{The surrogate model reduces the dimensions of the existing mTOV solutions and yet still keeps high accuracy.}
\MMG{The surrogate model is very quick}{This model is therefore very efficient} by a linear algebraic operation
rather than the iterative integration. \MG{}{The ROM-related techniques have been widely applied in GW science (see Ref.~\cite{Tiglio:2021ysj} for a review), e.g., fast evaluation of GW waveforms~\cite{Field2014PhRvX...4c1006F} and acceleration of GW parameter estimation~\cite{Canizares:2013ywa,Canizares:2014fya}.} Following the earlier work
of~\citet{Zhao2019PhysRevD.100.064034}, we extend our model to predict all the
PPK parameters, not just the orbital period decay parameter $\dot P_b$. To
explore EOS-dependent aspects, in this work, we choose 15 EOSs that are all
consistent with the maximum mass of NSs being larger than $2\,\mathrm{M_\odot}$.
This extends the number of EOSs in Ref.~\cite{Zhao2019PhysRevD.100.064034}.  We
use the central matter density $\rho_c$ of a NS to predict its radius $R$, mass
$m_A$, moment of inertia  $I_A$ and its derivative $k_A$, as well as the
effective scalar coupling $\alpha_A$ and its derivative $\beta_A$. Our models
\MG{are at}{keep} $\sim1\%$ level of accuracy.  According to our performance tests, one can
speed up the calculations by at least two and even three orders of magnitude for
the coupling parameters $\beta_A$ and $k_A$, and yet still keep the due
accuracy.  We demonstrate various applications with binary pulsars to illustrate
the practical value of our ROMs. 

The improvements of this work include the followings.
\begin{enumerate}[(I)]
    \item We use a larger set of EOSs.
    \item We calculate the mTOV equation to build ROMs in the DEF theory for
    slowly rotating NSs, instead of the nonrotating ones, to predict the moment
    of inertia and its derivative.
    \item We extend our ROMs to the coupling parameters by calculating the
    derivatives of $\alpha_A$ and $I_A$, and thus we can predict all the PPK
    parameters with new ROMs.
    \item Our ROMs speed up the calculation by two to even three orders of
    magnitude.
    \item We utilize the binary pulsars including double NSs to derive tight
    constraints on the DEF theory.
\end{enumerate}

The rest of this paper is organized as follows. Section~\ref{sec:theory} briefly
reviews the nonperturbative spontaneous-scalarization phenomena for slowly
rotating NSs\MG{. The}{ and discusses the} modifications of \MG{}{PPK parameters, including} orbital period decay, periastron advance
rate, and Einstein delay parameter in the DEF gravity \MG{are discussed}{}. In
\sect\ref{sec:method}, \MG{we analyze the difficulties in solving the mTOV equations
and calculating the derived parameters for large-scale calculations.}{we present the method of solving the mTOV equations, calculating the derived parameters, and constructing the ROMs for large-scale calculations.} \MG{We develop
a better numerical method and code it}{We code the model} in the \pystgromx\ (a.k.a.\
\texttt{pySTGROM eXtension}) package which is public for easy use for the
community.\footnote{\url{https://github.com/mh-guo/pySTGROMX}} In
\sect\ref{sec:constraint}, with \MG{the speedup from} \pystgromx, we \MG{constrain}{perform accelerated MCMC simulation and constrain} the
DEF theory tightly by combining the relevant PPK parameters available from
observations of five NS-white dwarf (WD) systems and three NS-NS systems.
\MG{Discussions and conclusions are given}{We summarize our conclusions} in \sect\ref{sec:conclusion}.

\section{Spontaneous scalarization in the DEF Theory}
\label{sec:theory}

The DEF theory is defined by the following general action in the {\it Einstein
frame}~\cite{Damour1993PhysRevLett.70.2220, Damour1996PhysRevD.54.1474}, 
\begin{eqnarray}
\label{eq:action}
S = &&\frac{c^4}{16\pi G_\star}\int\frac{\dd^4x}{c}\sqrt{-g_\star}\left[R_\star-2g^{\mu\nu}_\star\partial_\mu\varphi\partial_\nu\varphi-V(\varphi)\right]\nonumber\\
&&+S_m\left[\psi_m;A^2(\varphi)g^\star_{\mu\nu}\right] \, .
\end{eqnarray}
Here, $G_\star$ denotes the bare gravitational constant, $g_\star\equiv{\rm
det}\,g^\star_{\mu\nu}$ is the determinant of ``Einstein metric''
$g^\star_{\mu\nu}$, $R_\star$ is the Ricci curvature scalar of
$g^\star_{\mu\nu}$, and $\varphi$ is a dynamical scalar field. In the \MG{last term}{matter part $S_m$}
of \eq\eqref{eq:action}, $\psi_m$ denotes matter fields collectively, and \MG{the
conformal coupling factor $A(\varphi)$ describes how $\varphi$ couples to
$\psi_m$ in the Einstein frame}{it couples to
$\varphi$ by the
conformal coupling factor $A(\varphi)$}.  \MG{We}{In this study, we} assume that the potential, $V(\varphi)$, is
a slowly varying function at the typical scale of the system we consider and set
$V(\varphi)=0$ in our calculation for simplicity (see
Refs.~\cite{Ramazanoglu:2016kul, Xu:2020vbs} for a massive scalar field).

Varying the action~\eqref{eq:action} yields the field equations,
\begin{eqnarray}
\label{eq:field_eq}
R^\star_{\mu\nu}&=&\partial_\mu\varphi\partial_\nu\varphi + \frac{8\pi G_\star}{c^4}\left(T^\star_{\mu\nu} - \frac{1}{2}T^\star g^\star_{\mu\nu}\right)\,,\\
\label{eq:field_eq_scalar}
\Box_{g^\star}\varphi&=&-\frac{4\pi G_\star}{c^4} \alpha(\varphi)T_\star\,,
\end{eqnarray}
\MG{where $T^{\mu\nu}_\star\equiv2c(-g_\star)^{-1/2}\delta S_m / \delta
g^\star_{\mu\nu}$ denotes the matter stress-energy tensor,}{where the matter stress-energy tensor is
\begin{equation}
    T^{\mu\nu}_\star\equiv\frac{2c}{\sqrt{-g_\star}}\left(\frac{\delta S_m}{ \delta
    g^\star_{\mu\nu}}\right) \,,
\end{equation}
} and $T^\star\equiv
g^\star_{\mu\nu}T^{\mu\nu}_\star$ is the trace.  In
\eq(\ref{eq:field_eq_scalar}), the parameter $\alpha(\varphi)$ is defined as the
derivative of logarithmic $A(\varphi)$,
\begin{equation}
    \alpha(\varphi)\equiv\frac{\partial\ln A(\varphi)}{\partial\varphi} \,,
\end{equation}
which indicates the coupling strength between the scalar field and matters [see
Eq.~(\ref{eq:field_eq_scalar})].

In the DEF theory~\cite{Damour1996PhysRevD.54.1474}, $\ln A(\varphi)$ is
designated as
\begin{equation}
    \ln A(\varphi)=\frac{1}{2}\beta_0\varphi^2 \,\MG{.}{,}
\end{equation}
\MG{}{where
\begin{equation}
    \beta_0\equiv\frac{\partial^2\ln A(\varphi)}{\partial\varphi^2}\Bigg{|}_{\varphi=\varphi_0} \,,
\end{equation}
is a free parameter with $\varphi_0$ the asymptotic
scalar field value of $\varphi$ at spatial infinity. Then $\alpha(\varphi)=\partial\ln A(\varphi)/\partial\varphi=\beta_0\varphi$, and we further
denote $\alpha_0\equiv\beta_0\varphi_0$.} 
\MG{Then $\alpha(\varphi)=\partial\ln A(\varphi)/\partial\varphi=\beta_0\varphi$. We
denote $\alpha_0\equiv\beta_0\varphi_0$, where $\varphi_0$ is the asymptotic
scalar field value of $\varphi$ at spatial infinity.}{} Note that we have
$\alpha_0=\beta_0=0$ in GR.

For NSs, nonperturbative scalarization phenomena develop
when\MG{}{ $\beta_0\lesssim-4$}~\cite{Damour1993PhysRevLett.70.2220, Barausse2013PhysRevD.87.081506}\MG{}{.}
Generally, a more negative $\beta_0$ means more manifest spontaneous
scalarization in the strong-field regime.  In such case, the {\it effective
scalar coupling} for a NS ``$A$'' with a total mass-energy of $m_A$ is
\begin{equation}
    \alpha_A\equiv\frac{\partial\ln m_A(\varphi)}{\partial\varphi}\Bigg{|}_{\varphi=\varphi_0} \,,
\end{equation}
which measures the effective coupling strength between the scalar field and the NS.

Now we consider a scalarized NS in a binary system in the DEF theory. For a
binary pulsar system with the pulsar labeled ``$A$'' and its companion labeled
``$B$'', the parameters $\alpha_A$ and $\alpha_B$ contribute to the secular
change of the orbital period~\cite{Damour1996PhysRevD.54.1474}. In this work, we
investigate two contributions to $\dot{P}_b$, the dipolar contribution,
$\dot{P}_b^\mathrm{dipole}$, and the quadrupolar contribution,
$\dot{P}_b^\mathrm{quad}$. They are defined by~\cite{Damour1996PhysRevD.54.1474}
\begin{eqnarray}
\label{eq:dot_Pb_dipole}
\dot{P}_b^\mathrm{dipole}&=&-\frac{2\pi G_\star n_b}{c^3}g(e)\frac{m_Am_B}{m_A+m_B}\left(\alpha_A-\alpha_B\right)^2\,,\\
\label{eq:dot_Pb_quad}
\dot{P}_b^\mathrm{quad}&=&-\frac{192\pi G_{AB}^{5/3}n_b^{5/3}}{5c^5(1+\alpha_A\alpha_B)}f(e)\frac{m_Am_B}{(m_A+m_B)^{1/3}}\,,
\end{eqnarray}
where $n_b \equiv2\pi/P_b$, and
\begin{eqnarray}
g(e)\equiv&&\left(1-e^2\right)^{-5/2}\left(1+\frac{e^2}{2}\right)\,,\\
f(e)\equiv&&\left(1-e^2\right)^{-7/2}\left(1+\frac{73}{24}e^2+\frac{37}{94}e^4\right)\,.
\end{eqnarray}
Here the bare gravitational constant in \eq\eqref{eq:dot_Pb_dipole}, $G_\star$,
is obtained with the Newtonian constant $G_N$ by $G_\star
=G_N/(1+\alpha_0^2)$~\cite{Damour1992CQGra...9.2093D}, owing to the weak field
coupling. The body-dependent effective gravitational constant in
\eq\eqref{eq:dot_Pb_quad}, $G_{AB}$, is given by $G_{AB}\equiv G_\star
(1+\alpha_A\alpha_B)$.  The quadrupolar contribution $\dot{P}_b^\mathrm{quad}$
is close to the prediction of GR with a negligible correction, while the dipolar
contribution $\dot{P}_b^\mathrm{dipole}$ is the dominant additional contribution
in the DEF theory. Other subleading contributions induced by the scalar field
can be neglected in this study [see Eq.~(6.52) of
Ref.~\cite{Damour1992CQGra...9.2093D}].  Note that the effective scalar coupling
\MMG{is zero for BHs in the DEF theory due to the no-hair
theorem}{approaches to $\alpha_0$ for WDs in the weak field and becomes zero for BHs since the DEF theory still satisfies the no-hair theorem}~\cite{Abbott:2017vtc,Berti:2015itd}\MMG{ and approaches to $\alpha_0$ for WDs
in the weak field}{}. Thus, considering Eq.~\eqref{eq:dot_Pb_dipole}, the 
contribution of dipolar radiation to $\dot P_b$  plays an important role in
NS-WD, NS-BH, and asymmetric NS-NS binaries. In those binaries, there could be a
large difference between $\alpha_A$ and $\alpha_B$.

Similarly to $\alpha_A$, we define 
\begin{equation}\label{eq:betaA}
    \beta_A\equiv\frac{\partial\alpha_A}{\partial\varphi} \Bigg{|}_{\varphi=\varphi_0} \,,
\end{equation}
which is the strong-field analogue of the parameter $\beta_0$.  Then the
theoretical prediction for the periastron advance rate in the DEF theory
is~\cite{Damour1996PhysRevD.54.1474} 
\begin{eqnarray}
    \dot{\omega}^{\rm th}(m_A,m_B)&\equiv&\frac{3n_b }{1-e^2}\left(\frac{G_{AB}(m_A+m_B)n_b}{c^3}\right)^{2/3}\nonumber\\
    &&\times\left[\frac{1-\frac{1}{3}\alpha_A\alpha_B}{1+\alpha_A\alpha_B}-\frac{X_A\beta_B\alpha^2_A+X_B\beta_A\alpha^2_B}{6(1+\alpha_A\alpha_B)^2}\right]\,,\nonumber\\
\end{eqnarray}
where $X_A\equiv m_A/(m_A+m_B)\equiv1-X_B$.  Finally, we consider a
slowly-rotating NS with moment of inertia (in Einstein units) $I_A$.  We denote 
\begin{equation}\label{eq:kA}
    k_A\equiv-\frac{\partial\ln I_A}{\partial\varphi} \Bigg{|}_{\varphi=\varphi_0} \,,
\end{equation}
as the ``coupling factor'' for moment of
inertia~\cite{Damour1996PhysRevD.54.1474}.  The theoretical prediction of the
Einstein delay parameter in the DEF theory is~\cite{Damour1996PhysRevD.54.1474},
\begin{eqnarray}
    \gamma^{\rm th}(m_A,m_B)&\equiv&\frac{e}{n_b}\frac{X_B}{1+\alpha_A\alpha_B}\left(\frac{G_{AB}(m_A+m_B)n_b}{c^3}\right)^{2/3}\nonumber\\
    &&\times\left[X_B(1+\alpha_A\alpha_B)+1+K^B_A\right]\,,
\end{eqnarray}
where $K^B_A\equiv\alpha_B(m_B)k_A(m_A)$ describes the contribution from the
variation of $I_A$ under the influence of the companion $B$. 

For the (almost) symmetric double NS systems, such as the double pulsar PSR
J0737$-$3039~\cite{Kramer2006Sci...314...97K}, because of the similar binary
masses, $\alpha_A$ is very close to $\alpha_B$, leading to a tiny effect from
the dipolar radiation. Thus, it could be difficult to constrain the DEF theory
solely by the PPK parameter $\dot P_b$, unless $\dot P_b$ is extremely well
measured. For these
systems, instead of $\dot P_b$, the PPK parameters $\dot\omega$ and $\gamma$ can
be used of to provide better constraints, especially when both of the NSs
develop spontaneous scalarization.

\begin{figure}[t]
    \centering
    \includegraphics[width=0.98\linewidth]{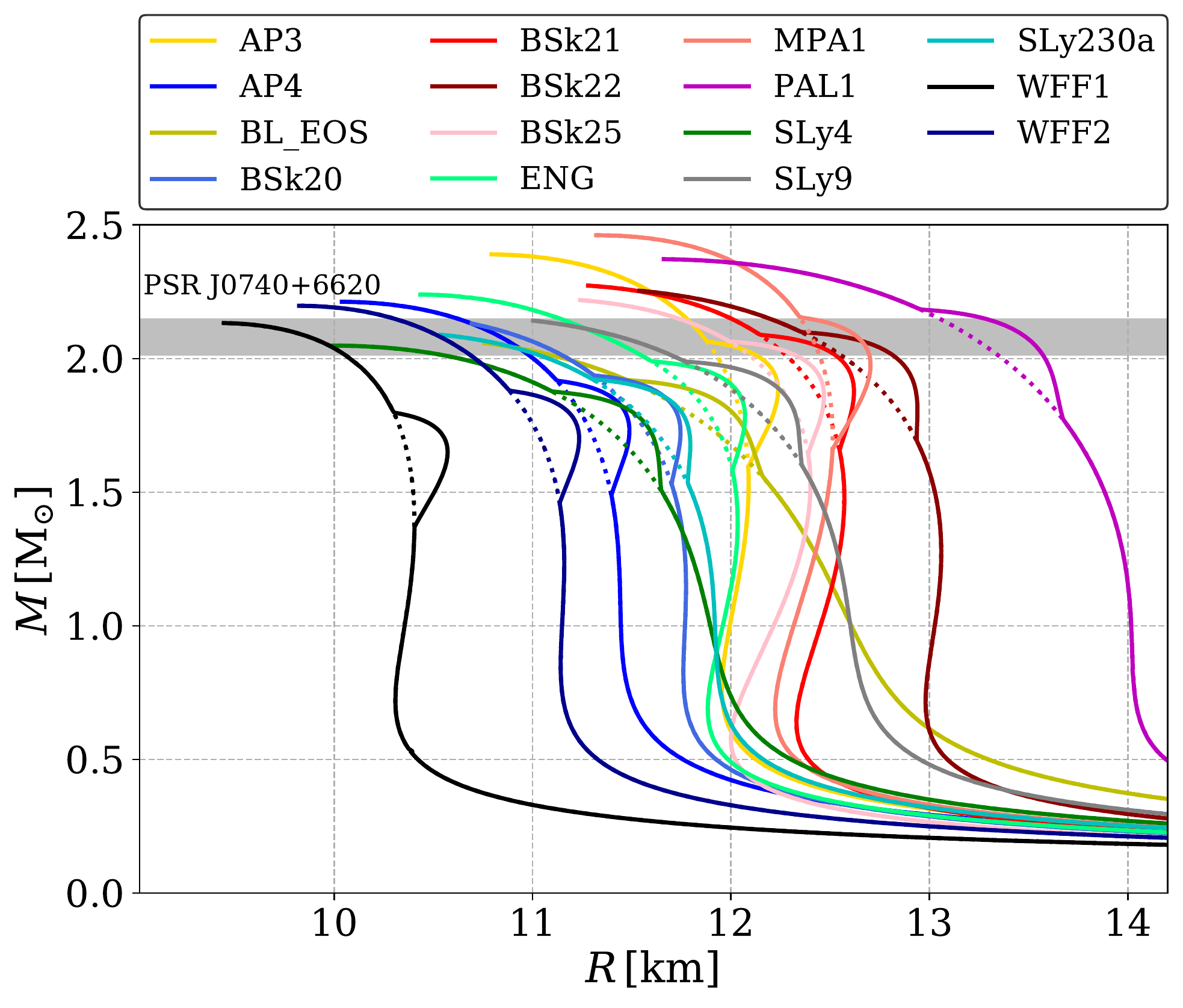}
    \caption{Mass-radius relations of NSs for the 15 EOSs that we adopt in this
    study. The mass-radius relations are derived from GR (dashed lines) and from
    a DEF theory with $\log_{10}|\alpha_0|=-5.0$ and $\beta_0=-4.5$ (solid
    lines).  The massive pulsar mass from PSR
    J0740$+$6620~\cite{Fonseca:2021wxt} is overlaid in gray.}
    \label{fig:MR}
\end{figure}
\begin{figure}[t]
    \centering
    \includegraphics[width=0.9\linewidth]{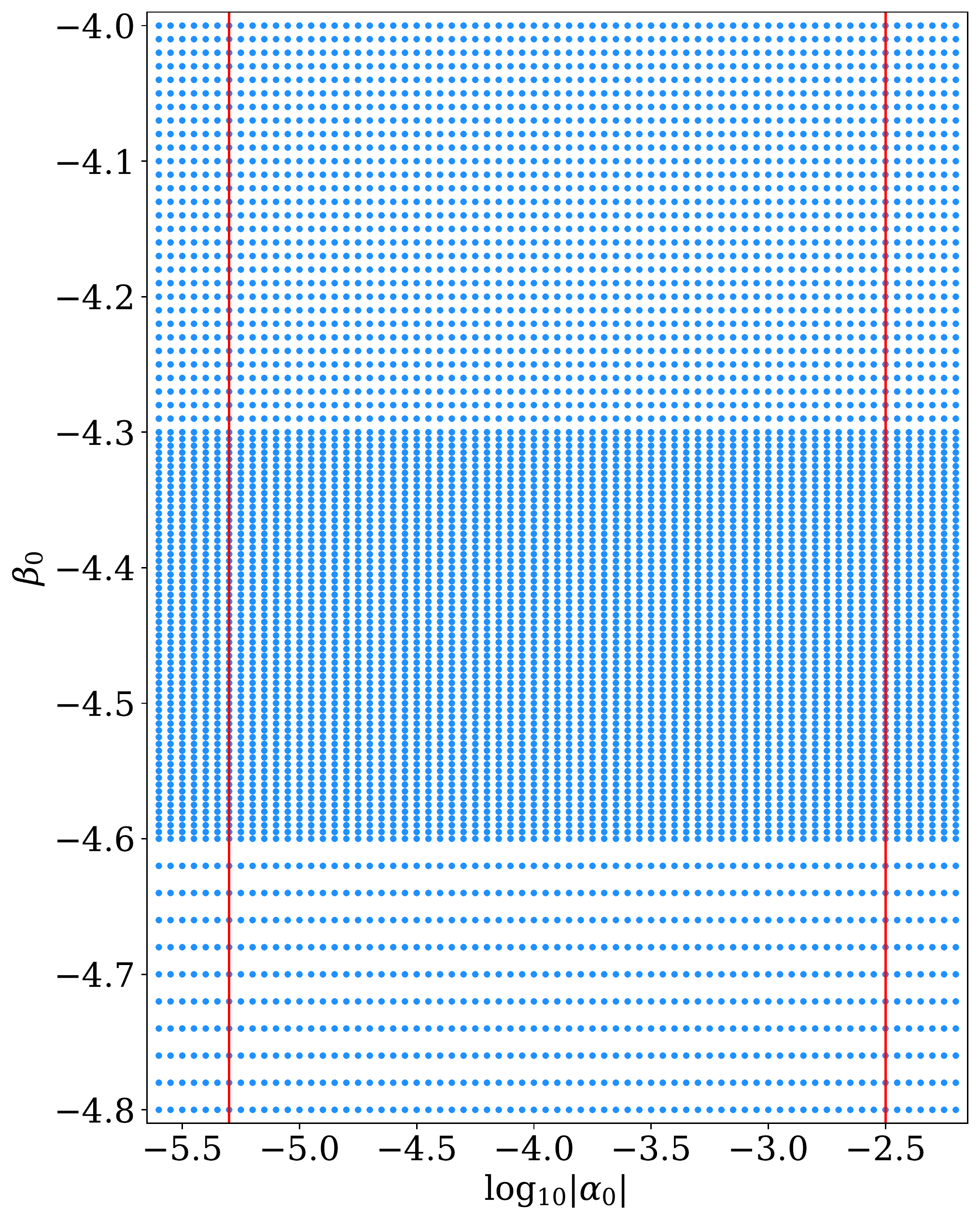}
    \caption{An uneven grid in the parameter space
    $\left(\log_{10}|\alpha_0|,-\beta_0\right)$ for calculating $\beta_A$ and
    $k_A$. We generate a set of $69\times101=6969$ parameter pairs as the
    training data in total. The region between red lines corresponds to the \MG{}{training} data
    we use for building ROMs.}
    \label{fig:grid}
\end{figure}
\begin{figure*}[t]
    \centering
    \includegraphics[width=0.9\linewidth]{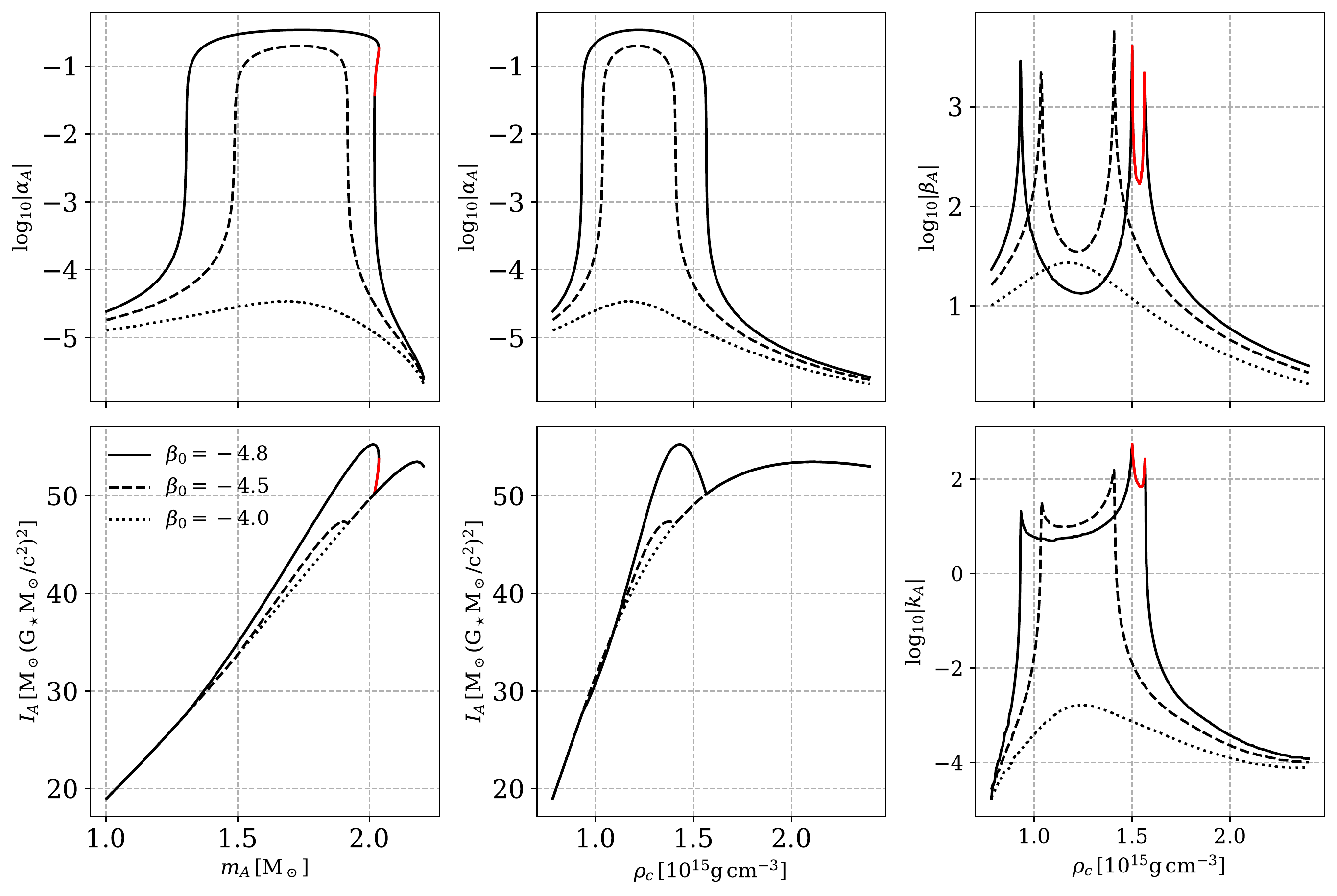}
    \caption{Illustration of spontaneous scalarization in the DEF gravity and
    the numerically pathological phenomena that occur when integrating the mTOV
    equations for the EOS AP4. \MG{The calculation assumes DEF theories with}{We adopt DEF parameter sets}
    $\log_{10}|\alpha_0|=-5.3$ and $\beta_0=-4.8$ (solid lines), $-4.5$ (dashed
    lines) and $-4.0$ (dotted lines). For $\log_{10}|\alpha_0|=-5.3$, the scalar
    field is weak for $\beta_0=-4.0$, strong for $\beta_0=-4.5$, and this causes
    remarkable pathological phenomena for $\beta_0=-4.8$. The red lines mark the
    pathological region. In this region, $\beta_A$ and $k_A$ are positive,
    instead of negative values as in most cases.  }
    \label{fig:curve}
\end{figure*}
\begin{figure}[t]
    \centering
    \includegraphics[width=0.95\linewidth]{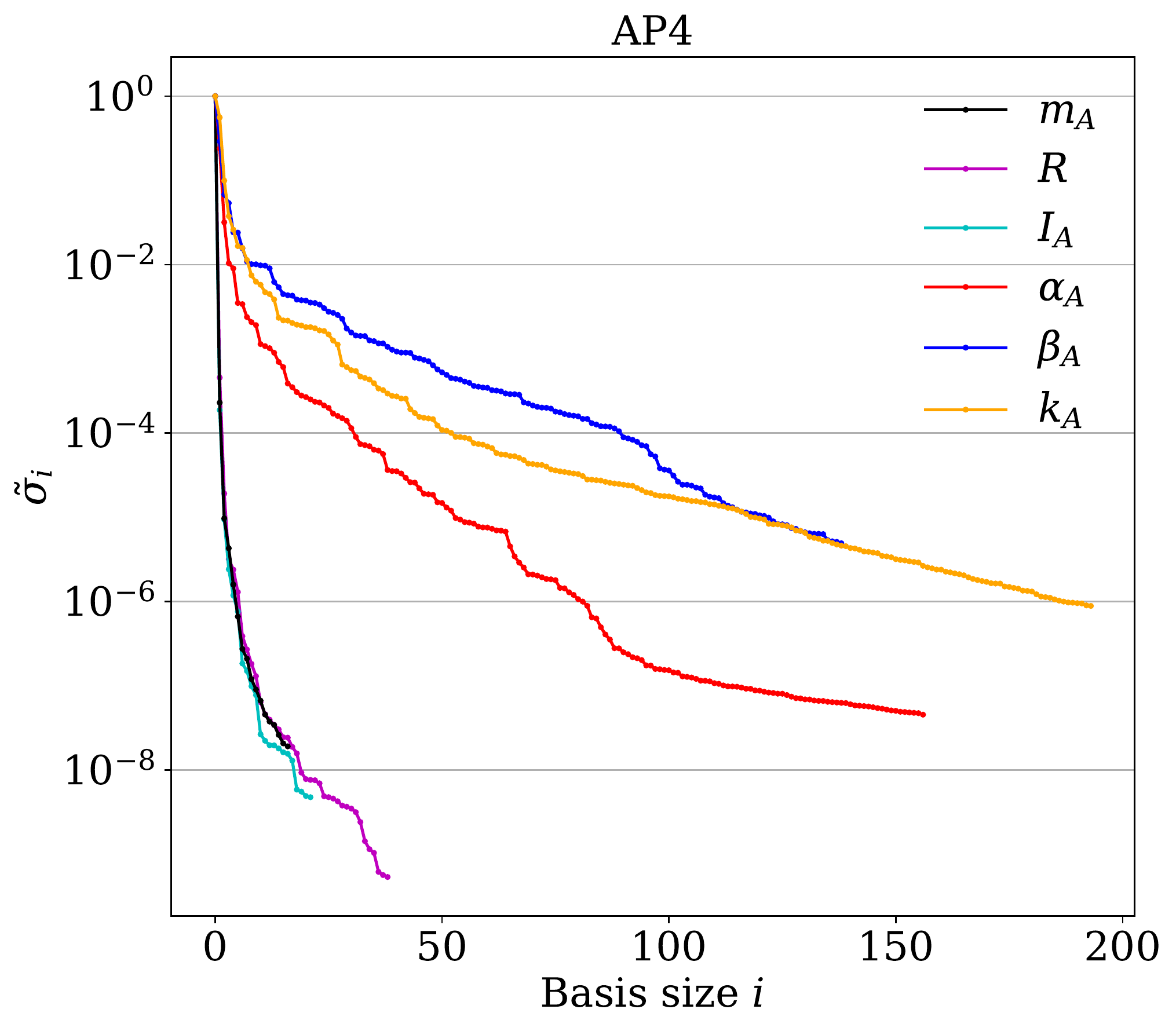}
    \caption{Relative maximum projection errors, $\tilde{\sigma}_i$, in \MG{building}{the process of constructing} 
    the ROMs for the EOS AP4. We set $\Sigma=10^{-7}$ for $m_A$, $R$, and $I_A$,
    $\Sigma=10^{-5}$ for $\alpha_A$, and $\Sigma=10^{-4}$ for $\beta_A$ and
    $k_A$. The $\tilde{\sigma}_i$'s of $m_A$, $R$ and $I_A$ decrease rapidly to
    $~10^{-8}$ with a basis size of $\sim 30$. On the contrary, the errors
    decline slowly for the coupling parameters $\alpha_A$, $\beta_A$, and $k_A$
    with roughly $100$--$200$ of basis size.}
    \label{fig:proj_err}
\end{figure}
\begin{figure}[t]
    \centering
    \includegraphics[width=0.95\linewidth]{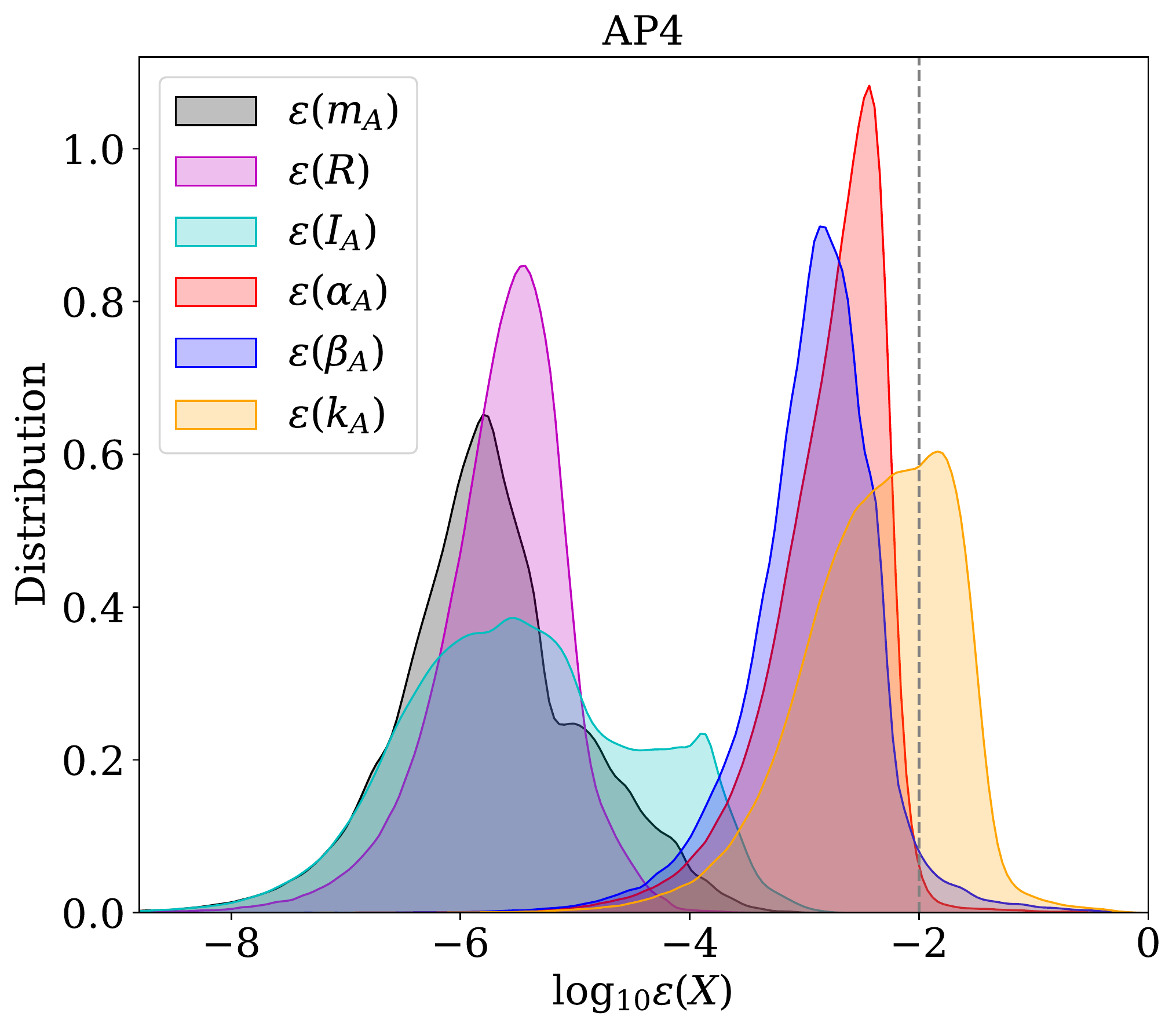}
    \caption{Kernel density estimation (KDE) distribution of the relative error
    $\varepsilon(X)$\MG{}{ defined in the text}, where $X\in\{m_A,R,I_A,\alpha_A,\beta_A,k_A\}$. The dashed
    line \MG{shows}{indicates} the relative tolerable error \MG{}{we set} in the mTOV integration
    ($\lesssim1\%$). The relative errors are small $(\lesssim10^{-4})$ for
    $m_A$, $R$, and $I_A$ but relatively large $(\lesssim10^{-2})$ for  the
    coupling parameters $\alpha_A$, $\beta_A$, and $k_A$. But the errors are
    negligible in most practical cases.}
    \label{fig:epsilon}
\end{figure}

\section{\label{sec:method} Extended ROM models}

\subsection{Solving the mTOV equations}\label{subsec:mTOV}

We here turn our attention to deriving the quantities of NSs in the strong
field.  For a specific EOS, given the initial conditions, namely the central
matter density $\rho_c$ and the central scalar field $\varphi_c$, one can
integrate the mTOV equations to obtain the solution of a slowly rotating NS. To
derive the prediction of a DEF theory (namely, with fixed $\alpha_0$,
$\beta_0$), one varies the initial condition $\varphi_c$ iteratively with the
``shooting method'' until the boundary solution matches the desired value of
$\varphi_0=\alpha_0/\beta_0$. Thus, given parameter set $\left(\alpha_0,\,
\beta_0,\, \rho_c \right)$, one can solve the mTOV equation by the shooting
method to obtain the macroscopic quantities of a NS (see
Ref.~\cite{Damour1996PhysRevD.54.1474} for details). The quantities contain the
NS radius $R$, the gravitational mass $m_A$, the baryonic mass $\bar{m}_A$, the
effective scalar coupling $\alpha_A$, and the moment of inertia $I_A$.  In this
way, we have a comprehensive description of the spontaneous scalarization.  In
\fig\ref{fig:MR}, we show mass-radius relation of NSs in a DEF theory with
$\log_{10}|\alpha_0|=-5.0$ and $\beta_0=-4.5$ for 15 EOSs we adopt in this
study. It indicates clearly that the spontaneous scalarization phenomena develop
for NSs with certain masses, and larger radii are \MG{}{usually} predicted in this range.

However, to determine the coupling parameters $\beta_A$ and $k_A$, we have to
calculate the derivatives with respect to the scalar field $\varphi_0$ from
Eqs.~\eqref{eq:betaA} and~\eqref{eq:kA} for a fixed form of the conformal
coupling factor $A(\varphi)$ (i.e., with a fixed $\beta_0$) and a fixed baryonic
mass $\bar{m}_A$. Calculations with different $\varphi_0$'s (or equivalently,
different $\alpha_0$'s) but the same $\beta_0$ and $\bar{m}_A$ are required. For
a single run, the calculation is generally performed by applying the shooting
method for both $\varphi_0$ and $\bar{m}_A$ with trial and
error~\cite{Damour1996PhysRevD.54.1474}. Therefore, obtaining $\beta_A$ and
$k_A$ is usually very time-consuming, especially for a large-scale calculation.
In this study, since we investigate the phenomena in the whole parameter space
of interest, we calculate the relevant derivatives on a grid covering the region
to avoid wasteful repetition.

In practice, for each EOS, we choose the range of $\rho_c$ so that
$m_A\in(1\,{\rm M_\odot},m^{\rm max}_A)$ with the EOS-dependent maximum NS mass
$m^{\rm max}_A$. We set an uneven spacing of $\rho_c$ in this range.  The
spacing of $\rho_c$ samples is smaller in the range where the coupling
parameters $\alpha_A$, $\beta_A$ and $k_A$ are rapidly changing. The number of
samples in $\rho_c$, namely $N_{\rho_c}$, varies from $\sim 300$ to $\sim 500$,
depending on the specific EOS.  Then we generate an uneven gird of
$\left(\log_{10}|\alpha_0|,\beta_0\right)\in[-5.6,-2.2]\times[-4.8,-4.0]$, as
shown in \fig\ref{fig:grid}.  The number of nodes in grid is set to
$N_{\alpha_0}\times N_{\beta_0}=69\times101=6969$.  We calculate the coupling
parameters $\beta_A$ and $k_A$ on each point with a reasonable differential
step.  Specifically, for each point $\left(\alpha_0,\beta_0,\rho_c\right)$, we
carefully select backward step $\Delta\varphi_0^{(1)}$ and forward step
$\Delta\varphi_0^{(2)}$. Thus, we have
$\varphi_0^{(1)}=\varphi_0-\Delta\varphi_0^{(1)}$ and
$\varphi_0^{(2)}=\varphi_0+\Delta\varphi_0^{(2)}$, or equivalently,
$\alpha_{0}^{(1)}$ and $\alpha_{0}^{(2)}$. Then we select points
$\left(\alpha_{0}^{(1)},\beta_0,\rho_c^{(1)}\right)$ and
$\left(\alpha_{0}^{(2)},\beta_0,\rho_c^{(2)}\right)$, where $\rho_c^{(1)}$ and
$\rho_c^{(2)}$ are chosen so that $\bar{m}_A$ can be fixed to the value of
$\bar{m}_A$ at the point $\left(\alpha_0,\beta_0,\rho_c\right)$.  By
interpolating the samples, we obtain the values $\alpha_A^{(1)}$,
$\alpha_A^{(2)}$, $I_A^{(1)}$, and $I_A^{(2)}$ at these two points.  Then we
calculate $\beta_A$ and $k_A$ numerically following
\begin{equation}
    \beta_A=\frac{\alpha_A^{(2)}-\alpha_A^{(1)}}{\varphi_{0}^{(2)}-\varphi_{0}^{(1)}} \,,
\end{equation}
and
\begin{equation}
    k_A= - \frac{\ln I_A^{(2)}-\ln I_A^{(1)}} {\varphi_{0}^{(2)}-\varphi_{0}^{(1)}} \,.
\end{equation}
Finally, we make use of the data of $\log_{10}|\alpha_0|\in[-5.3,-2.5]$, instead
of $\log_{10}|\alpha_0|\in[-5.6,-2.2]$, for the further calculation to avoid the
inaccuracy of derivatives at boundaries (see the region between the two red
lines in \fig\ref{fig:grid}).  The boundary value $\alpha_0\approx10^{-2.5}$ is
the upper limit given by the Cassini spacecraft~\cite{Bertotti:2003rm}, and
$\beta_0\lesssim-4.0$ corresponds to values where spontaneous scalarization
happens in the DEF theory. As a result, we have $N_{\alpha_0}'\times
N_{\beta_0}=57\times101=5757$ available nodes for constructing ROMs.

We have to point out that in practice it is difficult to calculate $k_A$ when
the scalar field is weak. In this case, a change in $I_A$ due to the weak scalar
field is comparable to the random noises induced by shooting method in solving
the mTOV equations. Therefore, it is hard to keep the calculation of $k_A$
accuracy.  Here we adopt an approximation that $k_A\propto\alpha_0$ when the
spontaneous scalarization is not excited~\cite{Damour1996PhysRevD.54.1474}.
Under this assumption, for a fixed $\beta_0$, we have $k_A\propto \varphi_0$ and
$\big(\ln I_A - \ln I_A^\mathrm{GR} \big) \propto \varphi_0^2$. Thus, we
choose a slightly large differential step and calculate
\begin{equation}
    k_A= - 2\varphi_{0}\frac{\ln I_A^{(2)}-\ln I_A^{(1)}} {\left(\varphi_{0}^{(2)}\right)^2-\left(\varphi_{0}^{(1)}\right)^2} \,,
\end{equation}
to reduce the influence of numerical noises. In addition, there are some
unexpected glitches in the results of derivatives. We remove these glitches and
perform interpolation with the nearby values instead.

\subsection{Constructing ROMs for the DEF theory}

To overcome the general time-consuming computation of the mTOV integration with
the shooting method, we build ROMs for the DEF theory to improve the
efficiency~\cite{Field2014PhRvX...4c1006F, Zhao2019PhysRevD.100.064034}.  In
brief, to \MG{generate}{construct} a ROM for a curve $h(t;\bm{\lambda})$ with \MMG{}{variable $t$ and} parameters
$\bm{\lambda}$, one \MG{provides}{performs a training for} a \MG{}{given} training space of data
$\mathbf{V}\equiv\{h(t;\bm{\lambda}_i)\}$ on a given grid of parameters. In this
space, one can define a ``special'' inner product $\langle
h(\cdot;\bm{\lambda}),h(\cdot;\bm{\lambda})\rangle$ with an inherited norm
$||h(\cdot;\bm{\lambda})||\equiv\langle h(\cdot;\bm{\lambda}),h(\cdot;\bm{\lambda})\rangle$.
Then one selects a certain number (denoted as $m$) of basis as a chosen space
$\mathbf{RV}=\{e_i\}^m_{i=1}$ with the reduced basis (RB) method. In practice,
given the starting RB $\left(i=0\right)$, \MG{one iteratively seeks for $m$
orthonormal RBs by iterating the Gram-Schmidt orthogonalization algorithm with
greedy selection to minimize the maximum projection error,}{one iteratively seeks for the $i$-th
orthonormal RB with
greedy selection to minimize the maximum projection error,}
\begin{equation}
    \sigma_i\equiv\max_{h\in\mathbf{V}}\left|\left|h(\cdot; \bm{\lambda})-\mathcal{P}_ih(\cdot;\bm{\lambda})\right|\right|^2\,,
\end{equation}
where $\mathcal{P}$ describes the projection of $h(t;\bm{\lambda})$ onto the
span of the first $i$ RBs. \MG{}{In each step, one applies Gram-Schmidt orthogonalization and normalization algorithm to avoid ill-conditioning of computation.} The process terminates when
$\sigma_{m-1}\lesssim\Sigma$, a user-specified error bound. Then every curve in
the training space is well approximated by
\begin{equation}
    h(t;\bm{\lambda})\approx\sum_{i=1}^{m}c_i(\bm{\lambda})e_i(t) \equiv \sum_{i=1}^m\big\langle h(\cdot;\bm{\lambda}),e_i(\cdot)\big\rangle e_i(t)\,,
\end{equation} 
where $c_i(\bm{\lambda})$ is the coefficient to be used for the ROM.  After the
RBs are built, one selects $m$ samples $\{t_i\}^m_{i=1}$ as empirical nodes with
empirical interpolation method~\cite{BARRAULT2004667}. Finally, at each
empirical node, one performs a fit \MMG{}{(particularly, a 2-dimensional 5th spline interpolation in this work\footnote{Other 2-dimensional fitting methods do not show practical difference.})} to the parameter space, $\{\bm{\lambda}_i\}$, 
and completes the construction of ROM. \MG{}{Given another set of parameters $(\tilde{t};\tilde{\lambda})$ within the boundary of $\mathbf{V}$, one can predict $\tilde{h}(\tilde{t};\tilde{\lambda})$ with the ROM.} More details can be found
in Ref.~\cite{Zhao2019PhysRevD.100.064034}.

Extending the work of~\citet{Zhao2019PhysRevD.100.064034}, we build ROMs for six
parameters, $R$, $m_A$, $I_A$, $\alpha_A$, $\beta_A$, and $k_A$, as functions of
the central matter density\MMG{ $\rho_c$}{, i.e., the curve variable $t=\rho_c$}, with specialized parameters $
\bm{\lambda}=\left(\alpha_0,\beta_0\right)$. We choose the implicit parameter
$\rho_c$ as an independent variable to avoid the multivalued relations between
$m_A$ and $R$, as well as $\alpha_A$ and
$I_A$~\cite{Zhao2019PhysRevD.100.064034}.  In \fig\ref{fig:curve}, we show the
relations of the parameters and illustrate the multivalued phenomena. \MG{}{A manifest spontaneous scalarization occurs for $\beta_0\lesssim-4.4$, as shown by the $\alpha_A$-$m_A$ curves.} When
$\beta_0\lesssim-4.6$, the $\alpha_A$-$m_A$ and $I_A$-$m_A$ curves are bent
backwards, leading to multiple values of $\alpha_A$ and $I_A$ for a given $m_A$.
This region is marked in red in \fig\ref{fig:curve}. The multivalued relations
vanish when we use the central matter density $\rho_c$ as an independent
variable. In the red region, $\beta_A$ and $k_A$ are not negative as normal, but
rather positive due to the multivalued relations.  Values of $\alpha_A$,
$\beta_A$, and $k_A$ span several orders of magnitude. In practice, we use
$\ln|I_A|$, $\ln|\alpha_A|$, $\ln|\beta_A|$, and $\ln|k_0 - k_A|$, instead of
$I_A$, $\alpha_A$, $\beta_A$, and $k_A$, for a better numerical performance,
where $k_0$ is an EOS-dependent constant set by hand to avoid negative values of
$-k_A$ in the weak scalar field regime (see Fig.~\ref{fig:curve}). We have
$k_0\lesssim0.2$ generally. The training space is set to 
\begin{equation}
    \mathbf{V}=\left\{\left(\log_{10}|\alpha_0|,\beta_0\right)\in[-5.3,-2.5]\times[-4.8,-4.0]\right\} \,,
\end{equation}
corresponding to the region between the two red lines shown in
\fig\ref{fig:grid}. As mentioned earlier, in total we have $N_{\alpha_0}'\times
N_{\beta_0}=57\times101=5757$ available nodes for building ROMs.

To balance the computation cost and the accuracy of ROMs, we set the error bound
$\Sigma=10^{-7}$ for $m_A$, $R$, and $I_A$, $\Sigma=10^{-5}$ for $\alpha_A$, and
$\Sigma=10^{-4}$ for $\beta_A$ and $k_A$.  The relative projection error
\MG{$\tilde{\sigma_i}\equiv \sigma_i/\sigma_0$}{$\tilde{\sigma}_i\equiv \sigma_i/\sigma_0$} as a function of the basis size is
shown in \fig\ref{fig:proj_err}. To achieve the desired projection error, the
basis size is $\sim 20$--$40$ for $m_A$, $R$, and $I_A$, but $\sim 150$--$200$
for $\alpha_A$, $\beta_A$, and $k_A$. It means that more RBs are essential to
keep the accuracy of $\alpha_A$, $\beta_A$, and $k_A$. Note that the training
space contains $\sim 6000$ curves.  It implies that, to achieve a certain
precision (as given by our $\Sigma$'s), the curves may exhibit redundancy in the
parameter space, i.e., the amount of information necessary to characterize the
relations in the DEF theory to a certain desired precision is smaller than
anticipated. Our ROM extracts $\sim1\%$ of the data but captures the essential
information and retains the original predictions of the DEF theory to sufficient
precision.  The precision loss in ROM building is negligible, considering the
tolerable error ($\sim 1\%$) involved in the shooting method and the calculation
of derivatives. This is also verified in assessing the accuracy of the ROMs
below. Eventually we perform the fits and complete the construction of the six
ROMs (i.e., for $R$, $m_A$, $I_A$, $\alpha_A$, $\beta_A$, and $k_A$) for each of
the 15 EOSs. We encapsulate those six ROMs for the DEF theory in the \pystgromx\
package for community use.

We \MG{test}{examine} the performance of the ROMs with randomly generated parameter sets
$(\alpha_0,\beta_0,\rho_c)$ on our \texttt{Intel Xeon E5-2697A V4} computers. We
find that the averaged time for generating the parameters improves from $\sim1$
second by solving the mTOV equations to $\sim 1$ millisecond (for $m_A$, $R$,
and $I_A$) and $\sim 3$ milliseconds (for $\alpha_A$, $\beta_A$, and $k_A$) by
linear algebraic operations in the ROMs. Note that traditionally calculating
$\beta_A$ and $k_A$ involves applying shooting method for both $\varphi_0$ and
$\bar{m}_A$. The time is usually tens of seconds for generating one point in
such calculation. Thus, our method improves the speed of calculating $\beta_A$
and $k_A$ by at least three orders of magnitude.

To assess the accuracy of the ROMs, we define \MG{}{a relative error}
\begin{equation}\label{eq:epsilon_X}
    \varepsilon(X)=\left| \frac{X_{\rm ROM}-X_{\rm mTOV}}{X_{\rm ROM}+X_{\rm mTOV}}\right| \,,
\end{equation}
where $X\in\{m_A,R,I_A,\alpha_A,\beta_A,k_A\}$, to indicate the fractional
accuracy of the ROMs. In \eq(\ref{eq:epsilon_X}), we denote $X_{\rm ROM}$ as the
prediction of ROM, and $X_{\rm mTOV}$ as the value from solving the mTOV
equations with the shooting algorithm.  The values $X_{\rm mTOV}$ in the test
space should also be calculated at different samples but with the same method to
avoid the extra error induced by the use of different methods.  Thus, instead of
randomly generating parameters, we set another grid as the test space which is
shifted from the training space for $\alpha_0$, $\beta_0$, and $\rho_c$, and
calculate the parameters in the same way. The test space has the sparser
distribution of $\beta_0$. Note that in this method we \MG{include}{consider} all errors for
our ROMs, including the \textit{interpolating} errors.

\MG{The distributions of $\varepsilon(X)$ are shown}{We show the distributions of $\varepsilon(X)$} in \fig\ref{fig:epsilon}. The
relative errors of $m_A$, $R$, and $I_A$ are \MG{}{sufficiently small to be} $\lesssim10^{-5}$. On the contrary,
relative errors of $\alpha_A$, $\beta_A$, and $k_A$ are mostly smaller than 1\%.
Although their errors are larger than those of $R$, $m_A$, and $I_A$, the errors
are, in most cases, still small enough to be neglected compared with the
numerical errors from the shooting method.  For $k_A$, in some cases the change
of $I_A$ is rather small, as we have mentioned in \sect\ref{subsec:mTOV},
leading to a relatively large numerical error when calculating the derivatives. 
Thus, a small fraction of prediction for $k_A$ in our ROM has a seemingly large
error in the range $\sim1$--$10\%$. However, we find that in such case we have
$k_A \ll 1$, thus the deviation in this region has little influence on our study
and will not have a practical effect.

\begin{figure}
    \centering
    \includegraphics[width=\linewidth]{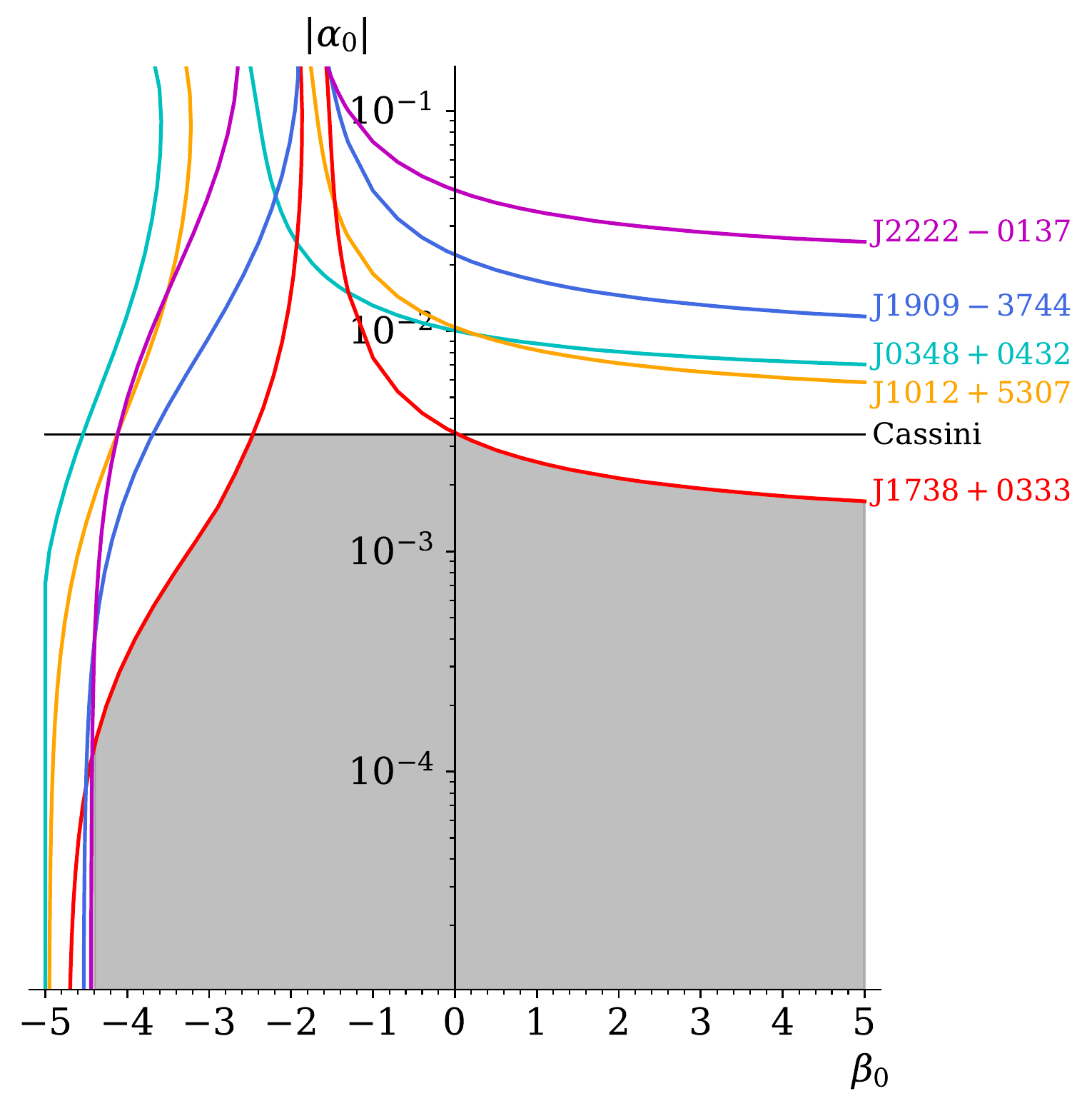}
    \caption{Constraints on the DEF parameters $\left(\alpha_0,\,\beta_0\right)$
    from a variety of binary pulsars on the DEF theory with the EOS AP4. Regions
    above \MG{}{the} curves are excluded, and the shaded area is the region that passes all
    tests in this plot. ``Cassini'' stands for the upper limit from the Shapiro
    time-delay measurement in the Solar System~\cite{Bertotti:2003rm}. \MG{}{Timing of binary pulsars provides constraints in strong field regime where $\beta_0\lesssim-4.0$.}}
    \label{fig:constraint}
\end{figure}
\begin{figure}
    \centering
    \includegraphics[width=\linewidth]{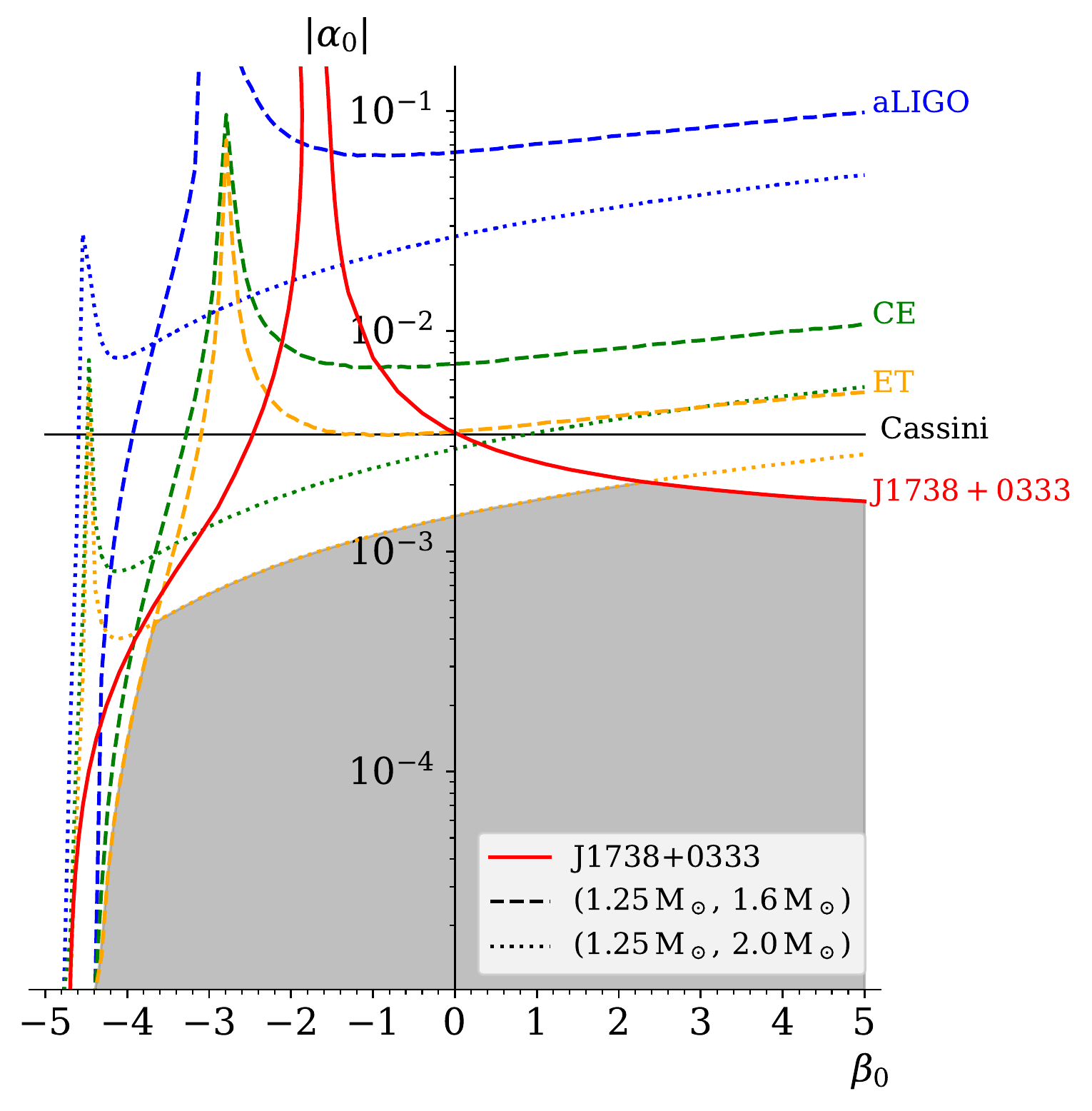}
    \caption{Similar to \fig\ref{fig:constraint}, but with {\it projected}
    limits from aLIGO, CE and ET with hypothetical GW signals of BNSs at a
    distance of 200\,Mpc. The dashed lines represent the constraints from a
    $\left(1.25\,\mathrm{M_\odot},\,1.6\,\mathrm{M_\odot}\right)$ BNS merger
    event, and the dotted lines correspond to a
    $\left(1.25\,\mathrm{M_\odot},\,2.0\,\mathrm{M_\odot}\right)$ event.  The
    shaded area is an illustration for combining all these constraints.  We plot
    the limits from PSR~J1738$+$0333 and the Cassini mission for comparison.}
    \label{fig:constraint_GW}
\end{figure}

\begin{table*}[ht]
    \caption{\label{tab:psrdat1}
    Binary parameters of the five NS-WD systems that we use to constrain the DEF
    theory (PSRs J0348$+$0432~\cite{Antoniadis:2013pzd},
    J1012$+$5307~\cite{Lazaridis2009MNRAS.400..805L,
    Desvignes2016MNRAS.458.3341D, Antoniadis2016arXiv160501665A,
    Sanchez2020MNRAS.494.4031M}, J1738$+$0333~\cite{Freire2012MNRAS.423.3328F},
    J1909$-$3744~\cite{Liu2020MNRAS.499.2276L}, and
    J2222$-$0137~\cite{Cognard2017ApJ...844..128C}).  The intrinsic time
    derivatives of the orbit period, $\dot{P}_b^{\rm int}$, are derived from
    $\dot{P}_b^{\rm obs}$ by subtracting the acceleration
    effect~\cite{Damour:1990wz} and other kinematic effects such as the
    ``Shklovskii'' effect~\cite{Shklovskii1970SvA....13..562S}. The Galactic
    acceleration effect is obtained from the latest Galactic potential model in
    Ref.~\cite{McMillan2017MNRAS.465...76M}.  For PSRs J0348$+$0432,
    J1012$+$5307, and J1738$+$0333, the pulsar masses, $m_A^\mathrm{obs}$, are
    obtained from the companion masses, $m_B^\mathrm{obs}$, and mass ratios,
    $q$. For PSRs J1909$-$3744 and J2222$-$0137, the masses are calculated from
    the Shapiro delay~\cite{Shao2017PhRvX...7d1025S}.  We list the standard
    $1$-$\sigma$ errors in units of the least significant digit(s) in
    parentheses.}
    \begin{ruledtabular}
    \begin{tabular}{llllll}
    Name &                                   J0348$+$0432 &        J1012$+$5307 &                                   J1738$+$0333 &         J1909$-$3744 &      J2222$-$0137 \\
    \colrule
    Orbital period, $P_b\,({\rm d})$       &                            0.102424062722(7) &  0.60467271355(3) &                          0.3547907398724(13) &  1.533449474305(5) &  2.44576454(18) \\
    Eccentricity, $e$        &                                 0.0000026(9) &      0.0000012(3) &                               0.00000034(11) &     0.000000115(7) &   0.00038096(4) \\
    Observed $\dot{P}_b$, $\dot{P}_b^{\rm obs}\,({\rm fs\,s^{-1}})$ &                                     $-273(45)$ &            50(14) &                                    $-17.0(31)$ &        $-510.87(13)$ &         200(90) \\
    Intrinsic $\dot{P}_b$, $\dot{P}_b^{\rm int}\,({\rm fs\,s^{-1}})$ &                                     $-274(45)$ &             $-5(9)$ &                                   $-27.72(64)$ &           $-4.4(79)$ &         $-60(90)$ \\
    Periastron advance, $\dot{\omega}\,({\rm deg\,yr^{-1}})$ &                                       --- &            --- &                                       --- &             --- &      0.1001(35) \\
    Einstein delay $\gamma\,({\rm ms})$    &                                       --- &            --- &                                       --- &             --- &          --- \\
    Pulsar mass, $m_A^\mathrm{obs}\,({\rm M_\odot})$    &                                      --- &            --- &                                       --- &          1.492(14) &         1.76(6) \\ 
    Companion mass, $m_B^\mathrm{obs}\,({\rm M_\odot})$    &  $0.1715_{-0.0030}^{+0.0045}$ &          0.174(7) &  $0.1817_{-0.0054}^{+0.0073}$ &           0.209(1) &       1.293(25) \\
    Mass ratio, $q\equiv m_A/m_B$        &                                    11.70(13) &           10.5(5) &                                       8.1(2) &             --- &          --- \\
    \end{tabular}
    \end{ruledtabular}
\end{table*}
\begin{table*}[ht]
    \caption{\label{tab:psrdat2}
    Binary parameters of the three NS-NS systems that we use to constrain the
    DEF theory (PSRs B1913$+$16~\cite{Weisberg2016ApJ...829...55W},
    J0737$-$3039A~\cite{Kramer2006Sci...314...97K}, and
    J1757$-$1854~\cite{Cameron2018MNRAS.475L..57C}). For PSR B1913$+$16, the
    masses are determined from measurement of periastron advance, $\dot\omega$,
    and Einstein delay, $\gamma$. For PSR J0737$-$3039A, the masses are
    determined from the measurement of $\dot\omega$ and theory-independent mass
    ratio $q$. For PSR J1757$-$1854, the masses are derived from DDGR model.
    Similar to \tab\ref{tab:psrdat1}, we list the standard $1$-$\sigma$ errors
    in units of the least significant digit(s) in parentheses.}
    \begin{ruledtabular}
    \begin{tabular}{llll}
    Name &           B1913$+$16 &       J0737$-$3039A &        J1757$-$1854 \\
    \colrule
    Orbital period, $P_b\,({\rm d})$       &  0.322997448918(3) &  0.10225156248(5) &  0.18353783587(5) \\
    Eccentricity, $e$        &       0.6171340(4) &      0.0877775(9) &     0.6058142(10) \\
    Observed $\dot{P}_b$, $\dot{P}_b^{\rm obs}\,({\rm fs\,s^{-1}})$ &           $-2423(1)$ &         $-1252(17)$ &        $-5300(200)$ \\
    Intrinsic $\dot{P}_b$, $\dot{P}_b^{\rm int}\,({\rm fs\,s^{-1}})$ &           $-2398(4)$ &         $-1252(17)$ &        $-5300(240)$ \\
    Periastron advance, $\dot{\omega}\,({\rm deg\,yr^{-1}})$ &        4.226585(4) &      16.89947(68) &        10.3651(2) \\
    Einstein delay $\gamma\,({\rm ms})$    &           4.307(4) &        0.3856(26) &         3.587(12) \\
    Pulsar mass, $m_A^\mathrm{obs}\,({\rm M_\odot})$    &           1.438(1) &         1.3381(7) &         1.3384(9) \\
    Companion mass, $m_B^\mathrm{obs}\,({\rm M_\odot})$    &           1.390(1) &         1.2489(7) &         1.3946(9) \\
    Mass ratio, $q\equiv m_A/m_B$        &             --- &            1.0714(11) &            --- \\
    \end{tabular}
    \end{ruledtabular}
\end{table*}

\section{\label{sec:constraint}Constraints from binary pulsars} 

Constraints on the DEF parameters, $\alpha_0$ and $\beta_0$, have been
investigated from various gravitational systems.  In the Solar System, the
Cassini mission~\cite{Bertotti:2003rm} \MG{gives a limit}{places a constraint} on $\alpha_0$,
$|\alpha_0|<3.4\times10^{-3}$ at \MG{68\% confidence level (CL)}{at $1$-$\sigma$ uncertainty}, from the
measurement of Shapiro delay in the weak-field regime. In the strong-field
regime, timing of binary pulsars provides the most stringent constraints so far
to the DEF theory~\cite{Damour1996PhysRevD.54.1474, Antoniadis:2013pzd,
Freire2012MNRAS.423.3328F, Shao2017PhRvX...7d1025S, Anderson:2019eay}.  Here we
extend the analysis in Refs.~\cite{Shao2017PhRvX...7d1025S,
Zhao2019PhysRevD.100.064034}, by carefully selecting more systems and
constraining the DEF parameters in a numerically faster and reliable way with
the ROMs we built in \sect\ref{sec:method}.  We \MG{will}{} employ the MCMC
technique~\cite{ForemanMackey:2012ig, Shao2017PhRvX...7d1025S} to derive
constraints \MG{when}{by} combining the timing observations from binary pulsar systems,
including NS-WD and NS-NS systems. We \MG{will}{} also discuss the improvements on
deriving NS properties and the constraints on the DEF parameters.

\subsection{Setup}

Previous work~\cite{Shao2017PhRvX...7d1025S,Zhao2019PhysRevD.100.064034} has
constrained the spontaneous scalarization of the DEF theory via MCMC simulations
by combining \MG{}{observations of the PPK parameter $\dot{P}_b$ of} some well-timed binary pulsars.  These previous studies considered
five well-timed NS-WD binary pulsars, PSRs
J0348$+$0432~\cite{Antoniadis:2013pzd},
J1012$+$5307~\cite{Lazaridis2009MNRAS.400..805L, Desvignes2016MNRAS.458.3341D,
Antoniadis2016arXiv160501665A, Sanchez2020MNRAS.494.4031M},
J1738$+$0333~\cite{Freire2012MNRAS.423.3328F},
J1909$-$3744~\cite{Liu2020MNRAS.499.2276L}, and
J2222$-$0137~\cite{Cognard2017ApJ...844..128C}.  We include these five systems
in our study as well and their latest relevant parameters are listed in
\tab\ref{tab:psrdat1}. For these systems, WD companions are weakly
self-gravitating objects, leading to a tiny effective scalar coupling
$\alpha_B\approx\alpha_0$. It indicates a large dipolar contribution
$\propto(\alpha_A-\alpha_0)^2$ to $\dot{P}_b$, if $\alpha_A$ is sufficiently
different from $\alpha_0$  [see \eq\eqref{eq:dot_Pb_dipole}].  As explained in
the caption of \tab\ref{tab:psrdat1}, the measurements of the masses are
independent of the PPK parameter $\dot{P}_b$.  Thus, we can safely place bounds
on the parameters of the DEF theory by utilizing these masses and the intrinsic
orbital decay rate, $\dot{P}_b^\mathrm{int}$.

Apart from the five NS-WD systems, we add three NS-NS systems in our study. As
discussed above, due to the closeness of $\alpha_A$ and $\alpha_B$, the dipolar
radiation of these (almost) symmetric double NS systems is small, making it
difficult to constrain the DEF theory with $\dot{P}_b$.  However, as shown in
\sect\ref{sec:theory}, some other PPK parameters derived from pulsar timing,
such as  the periastron advance rate $\dot{\omega}$ and the Einstein delay
parameter $\gamma$, are also modified in the DEF theory.  These modifications are
considerable, especially for NS-NS binaries.  Therefore, these double NS
systems are potentially powerful laboratories to constrain the free parameters
of the DEF theory.  In \tab\ref{tab:psrdat2}, we show parameters of three
well-timed double NSs used in this study: PSRs
B1913$+$16~\cite{Weisberg2016ApJ...829...55W},
J0737$-$3039A~\cite{Kramer2006Sci...314...97K}, and
J1757$-$1854~\cite{Cameron2018MNRAS.475L..57C}. In these systems, the
measurements of the masses depend on PPK parameters. For PSR B1913$+$16, the
masses are determined from measurement of periastron advance, $\dot\omega$, and
Einstein delay, $\gamma$. For PSR J0737$-$3039, the masses are determined from
the measurement of $\dot\omega$ and theory-independent mass ratio $q$. For PSR
J1757$-$1854, the masses are determined from the DDGR
model~\cite{1987grg..conf..209T,Taylor:1989sw}, which includes the contribution
from all PPK parameters via their GR formats.  In principle, only the
independent subset of  PPK parameters that were not used in deriving masses can
be utilized to constrain the DEF parameters.

To obtain NSs' properties, we need to assume a particular EOS. In this study, as
mentioned earlier, we adopt 15 EOSs, namely AP3, AP4, BL\_EOS, BSk20, BSk21,
BSk22, BSk25, ENG, MPA1, PAL1, SLy4, SLy9, SLy230a, WFF1, and WFF2 (see
Ref.~\cite{Lattimer2012ARNPS..62..485L} for a review).  The mass-radius
relations from these EOSs are shown in \fig\ref{fig:MR}.  These EOSs are all
consistent with the observed $2\,\mathrm{M_\odot}$ maximum mass limit of NSs,
and they are chosen inclined towards those which predict NS \MMG{radii}{radius} around
$\sim11$--$13\,\mathrm{km}$ \MMG{}{for a mass of $1.4 \, \mathrm{M_\odot}$}, as favored by the tidal deformability measurement
in GW170817~\cite{TheLIGOScientific:2017qsa, De:2018uhw, Capano:2019eae}. 

\subsection{Constraints from individual pulsars}

Before the MCMC simulations, we can estimate the constraints on the DEF
parameters, $\alpha_0$ and $\beta_0$, from individual binary systems. We
saturate the uncertainties on individual PPK parameters to determine the upper
limits of the DEF parameters. By determining if the predicted value
$\dot{P}_b^\mathrm{th}$ from \eqs\eqref{eq:dot_Pb_dipole} and
\eqref{eq:dot_Pb_quad} lies within the \MG{$1$-$\sigma$}{$2$-$\sigma$} range of $\dot{P}_b^{\rm
int}$, i.e., \MG{$\dot{P}_b^{\rm int}\pm\delta\dot{P}_b^{\rm int}$}{$\dot{P}_b^{\rm int}\pm2\delta\dot{P}_b^{\rm int}$}, we obtain the
limits on the difference of  effective scalar couplings $\alpha_A$ and
$\alpha_B$, i.e.,  $|\Delta\alpha|=\left|\alpha_A-\alpha_B\right|$. With it, we
constrain parameters $\alpha_0$ and $\beta_0$. 

In \fig\ref{fig:constraint}, we show the constraints on $\alpha_0$ and $\beta_0$
obtained from binary pulsars. In these systems, PSR J1738$+$0333 gives almost
the most stringent constraint. The upper limit of $\beta_0$ can reach $\sim
-4.4$ from PSRs J1012+5307 and J2222$-$0137.  When $\beta_0$ is around $-3$ to
$-2$, the constraints from binary pulsars are weak, leading to a peak of
$|\alpha_0|$. However, the limit from the Cassini mission helps exclude this
region. In \fig\ref{fig:constraint_GW}, we show constraints from hypothetical
asymmetric BNS merger events at 200\,Mpc, to be probed by GW detectors including
Advanced LIGO (aLIGO), CE, and ET~\cite{Shao2017PhRvX...7d1025S}. We assume two
hypothetical BNSs with masses, $(1.25\,\mathrm{M_\odot},\,
1.6\,\mathrm{M_\odot})$ and $(1.25\,\mathrm{M_\odot},\, 2.0\,\mathrm{M_\odot})$,
and the limits are obtained from Fisher information matrix~\cite{Finn:1992wt}.
It is shown that the GW detectors could help in constraining the DEF parameters
if such BNSs are observed.  Among the three detectors, ET would probably provide
tighter bounds, due to its lower starting frequency.  Note that these
constraints are dependent on the specific EOS, and in Figs.~\ref{fig:constraint}
and \ref{fig:constraint_GW} we have assumed  EOS AP4. The results are similar to
previous results in Refs.~\cite{Freire2012MNRAS.423.3328F,
Shao2017PhRvX...7d1025S}.  

\subsection{The Bayesian inference and the MCMC framework}

We here explore the constraints on the DEF theory combining the well-timed
binary pulsars through MCMC simulations.  MCMC technique provides a convenient
and statistically sound algorithm to generate the distribution of unknown
parameters for a model with many free parameters.  In this study, we use
\texttt{EMCEE},\footnote{\url{https://github.com/dfm/emcee}} a python
implementation of an affine-invariant MCMC ensemble sampler, and our \pystgromx\
package to speed up the calculation within the Bayesian framework.

In the Bayesian inference, given priors, hypothesis $\mathcal{H}$, data
$\mathcal{D}$, and all the extra relevant information $\mathcal{I}$, the
posterior distribution of the DEF parameters ($\alpha_0$, $\beta_0$),
$P(\alpha_0,\beta_0|\mathcal{D},\mathcal{H},\mathcal{I})$, can be inferred by
\begin{eqnarray}
    \label{eq:Bayes}
    P(\alpha_0,\beta_0|\mathcal{D}&&,\mathcal{H},\mathcal{I})=\nonumber\\
    &&\int\frac{P(\mathcal{D}|\alpha_0,\beta_0, \bm{\Xi},\mathcal{H},\mathcal{I})P(\alpha_0,\beta_0,\bm{\Xi}|\mathcal{H},\mathcal{I})}{P(\mathcal{D}|\mathcal{H},\mathcal{I})}\nonumber\dd \bm{\Xi} \,,\\    
\end{eqnarray}
where $\bm{\Xi}$ denotes all the \MG{other}{extra} unknown parameters in the theory. In
\eq\eqref{eq:Bayes},
$P(\mathcal{D}|\alpha_0,\beta_0,\bm{\Xi},\mathcal{H},\mathcal{I})$ is the
likelihood function, $P(\alpha_0,\beta_0,\bm{\Xi}|\mathcal{H},\mathcal{I})$ the
prior, and 
\begin{equation}
    P(\mathcal{D}|\mathcal{H},\mathcal{I}) = \int P(\mathcal{D}|\bm{\Xi},\mathcal{H},\mathcal{I}) P(\bm{\Xi}|\mathcal{H},\mathcal{I}) \dd\bm{\Xi} \,,
\end{equation}
the \textit{model evidence} which merely plays a role of normalization
here~\cite{DelPozzo:2016ugt}.  Based on \eq\eqref{eq:Bayes}, MCMC method
generates a sampling of values $(\alpha_0,\,\beta_0)$ that satisfies the
posterior distribution.

In the MCMC simulations, the parameters, $\alpha_0$ and $\beta_0$, are
constrained by evaluating the likelihood function. Considering all the binary
pulsar systems we include, we use a general logarithmic likelihood function,
\begin{eqnarray}
    \ln\mathcal{L}_\mathrm{PSR}&=&
    -\frac{1}{2}\sum_{i=1}^{N_\mathrm{PSR}}\left[
    \left(\frac{\dot{P}_b-\dot{P}_b^\mathrm{int}}{\sigma_{\dot{P}_b^\mathrm{int}}}\right)^2+
    \left(\frac{\dot{\omega}-\dot{\omega}^\mathrm{obs}}{\sigma_{\dot{\omega}^\mathrm{obs}}}\right)^2+
    \left(\frac{\gamma-\gamma^\mathrm{obs}}{\sigma_{\gamma^\mathrm{obs}}}\right)^2
    \nonumber \right. \\ 
    && \left. +
    \left(\frac{m_A-m_A^\mathrm{obs}}{\sigma_{m_A^\mathrm{obs}}}\right)^2+
    \left(\frac{m_B-m_B^\mathrm{obs}}{\sigma_{m_B^\mathrm{obs}}}\right)^2
    \right]_i \,,
    \label{eqn:log-likeli}
\end{eqnarray}
for $N_\mathrm{PSR}$ binary pulsars. Here we assume that observations of
different binary pulsars are independent.  The intrinsic orbital decay
$\dot{P}_b^\mathrm{int}$, the periastron advance rate $\dot\omega^\mathrm{obs}$,
the Einstein delay parameter $\gamma^\mathrm{obs}$, the pulsar mass
$m_A^\mathrm{obs}$, and the companion mass $m_B^\mathrm{obs}$ are given in
\tabs\ref{tab:psrdat1} and \ref{tab:psrdat2}.  \MG{The $1$-$\sigma$ uncertainty,
$\sigma_X$, is the observational uncertainty for parameter $X$, where
$X\in\left\{\dot{P}_b^\mathrm{int}, \dot\omega^\mathrm{obs},
\gamma^\mathrm{obs}, m_A^\mathrm{obs}, m_B^\mathrm{obs}\right\}$.}{Their observational $1$-$\sigma$ uncertainties are denoted as $\sigma_X$, where
$X\in\left\{\dot{P}_b^\mathrm{int}, \dot\omega^\mathrm{obs},
\gamma^\mathrm{obs}, m_A^\mathrm{obs}, m_B^\mathrm{obs}\right\}$.}  The
parameters, $\dot{P}_b$, $\dot\omega$, $\gamma$, and $m_A$ are dependent on the
parameter set, $(\alpha_0,\beta_0,\bm{\Xi})$. The companion mass $m_B$ is
consistently given from the parameter set $(\alpha_0,\beta_0,\bm{\Xi})$ as well
when it is a NS.  Otherwise, for a WD, it is \MG{picked}{generated} randomly within the
$1$-$\sigma$ uncertainty of $m_B^\mathrm{obs}$.

For WD companions, the coupling parameters can be reduced to $\alpha_B=\alpha_0$
and $\beta_B=\beta_0$. Thus, their contributions to $\dot\omega$ and $\gamma$
can be neglected.  In addition to the five NS-WD binary pulsars that have been
investigated in Refs.~\cite{Shao2017PhRvX...7d1025S,
Zhao2019PhysRevD.100.064034}, we combine three extra double NS systems. In these
NS-NS systems, we could utilize the information from $\dot{\omega}$ and
$\gamma$.  However, not each pulsar's measurements of $\dot{\omega}$, $\gamma$,
$m_A$, and $m_B$ are independent, as we have mentioned above. Thus, only for some
suitable pulsar systems, the contributions of $\dot{\omega}$ and $\gamma$ are counted
in a consistent way. In this study, we combine the contributions from
$\dot{P}_b^\mathrm{int}$ of eight binary pulsars and $\gamma$ of PSR
J0737$-$3039A in total for our estimation. In this way, we utilize the available
information from the timing data as much as possible.

Note that the other parameters included in MCMC calculations, such as the
orbital period, $P_b$, and the orbital eccentricity, $e$, are \MG{determined very
well}{well-measured} from the observations (see \tabs\ref{tab:psrdat1} and \ref{tab:psrdat2}).
Thus, \MG{we}{it is sufficient to} adopt their central values for simplicity in the process of MCMC
calculations\MG{, which is sufficient in the context}{}.

\MG{Now, we describe the details of employing the MCMC technique to get the
posteriors from the priors on $\left(\alpha_0,\beta_0\right)$ and the
log-likelihood function~\eqref{eqn:log-likeli}.}{} In the MCMC calculation, the
information we utilize to constrain the parameter space of
$\left(\alpha_0,\beta_0\right)$ is from the NSs. In total, we have
$N=N_\mathrm{NS-WD}+2N_\mathrm{NS-NS}$ NSs from the observations. To describe
the observational signatures fully in the DEF theory, we need $N+2$ free
parameters, collectively denoted as $\boldsymbol{\theta}=\left\{\alpha_0,\,
\beta_0,\, \rho_c^{(i)}\,(i=1,2,...,N)\right\}$ where $\rho_c^{(i)}$ is the
central matter density of NS $i$ in the Jordan frame.  The initial values of
$\rho_c^{(i)}$ are sampled around their GR values. We allow the simulation to
explore a large region of central matter density. Thanks to our ROMs, the
initial central values of the scalar field $\varphi_c$ are no longer needed
during the calculation, which avoids extra computationally intensive
calculations~\cite{Shao2017PhRvX...7d1025S}.  Using $\alpha_0$, $\beta_0$, and
$\rho_c^{(i)}$, we obtain the properties of NSs. Then we calculate the PPK
parameters and the log-likelihood functions to evaluate the posteriors. 

\MG{For our studies,}{In this study,} we carefully choose the priors of $(\alpha_0,\,\beta_0)$ to
cover the region of interest where the spontaneous scalarization develops. For
each MCMC run, we assume a uniform prior distribution of
$\left(\log_{10}|\alpha_0|,\, \beta_0\right)$ in the region of our ROMs, i.e.,
$\left(\log_{10}|\alpha_0|,\,\beta_0 \right)\in[-5.3,-2.5]\times[-4.8,-4.0]$.  

For each MCMC simulation, we use 32 walkers (chains) and 100,000 steps for each
walker. Thus, we produce 3.2 million samples in total.  \MG{The first half chain
samples are discarded as the \textit{BURN-IN}
phase}{The last half chain samples are remained and ``thined'', with a thinning factor of ten, to reduce dependence on the initial values and correlation of the
adjacent samples}~\cite{ForemanMackey:2012ig,Gelman:1992zz}. \MG{Then we ``thin'' the remaining
samples, with a thinning factor of ten, to reduce the correlation of the
adjacent samples.}{}
We perform the Gelman-Rubin test for convergence, and we have
confirmed that our samples pass the test~\cite{Gelman:1992zz}. Thus, our
posteriors of $\alpha_0$ and $\beta_0$ are statistically reliable.  We end up
with $1.6\times 10^5$ ``thinned'' samples for each simulation to infer the
marginalized distributions of the free parameters $\alpha_0$ and $\beta_0$.
Eventually, we apply this procedure to all the 15 EOSs.

\begin{figure}
    \centering
    \includegraphics[width=0.95\linewidth]{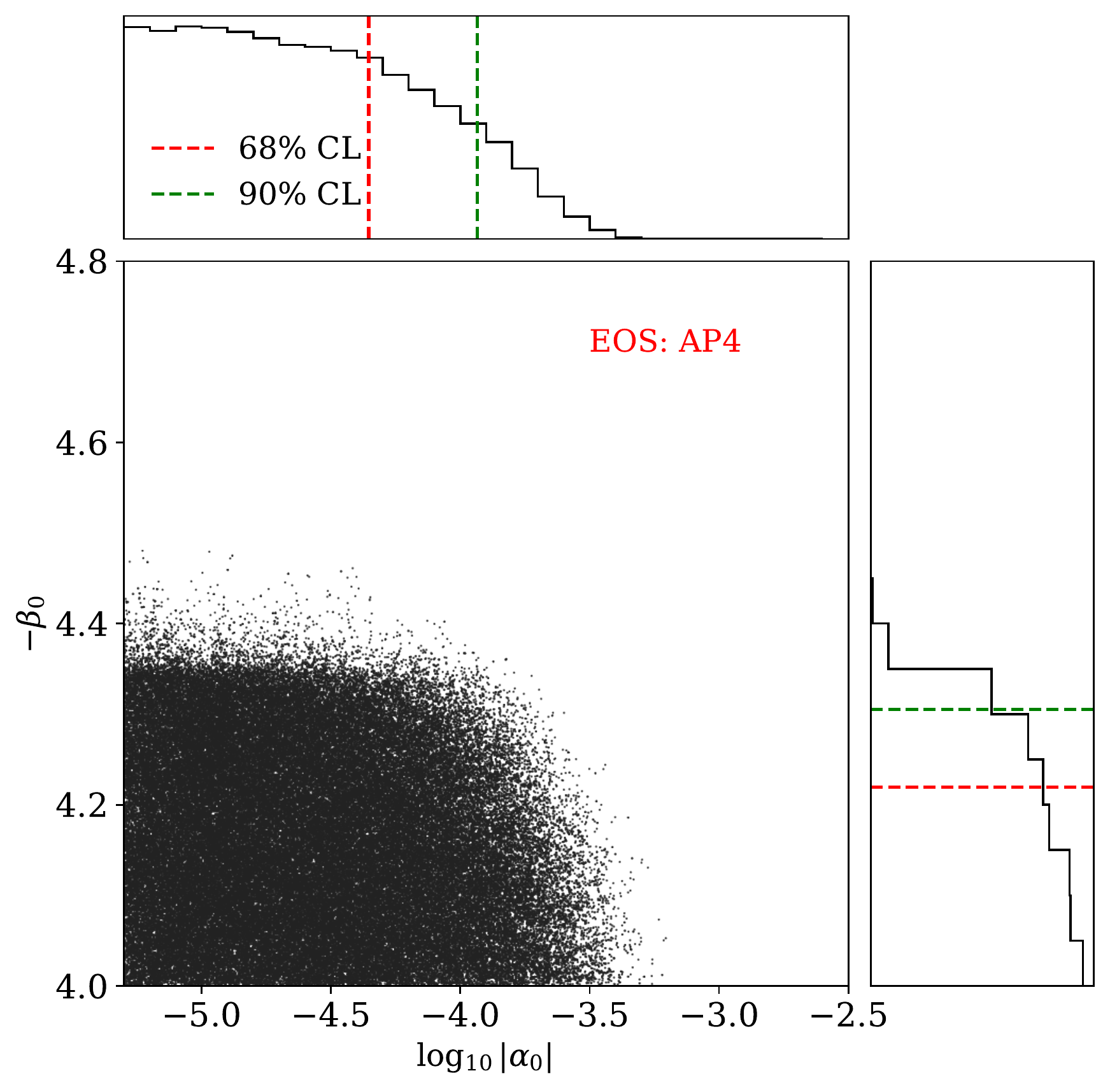}
    \caption{\MG{The marginalized 2-d}{Marginalized 2-d posteriors} distribution \MG{in the parameter space of}{for}
    $\left(\log_{10}|\alpha_0|,-\beta_0\right)$ from MCMC runs on the eight
    pulsars listed in \tabs\ref{tab:psrdat1} and \ref{tab:psrdat2} \MG{for}{with} the EOS
    AP4. The marginalized 1-d distributions and the corresponding upper limits
    at 68\% CL and 90\% CL are illustrated in upper and right panels.}
    \label{fig:mcmc}
\end{figure}
\begin{figure*}
    \centering
    \includegraphics[width=0.8\linewidth]{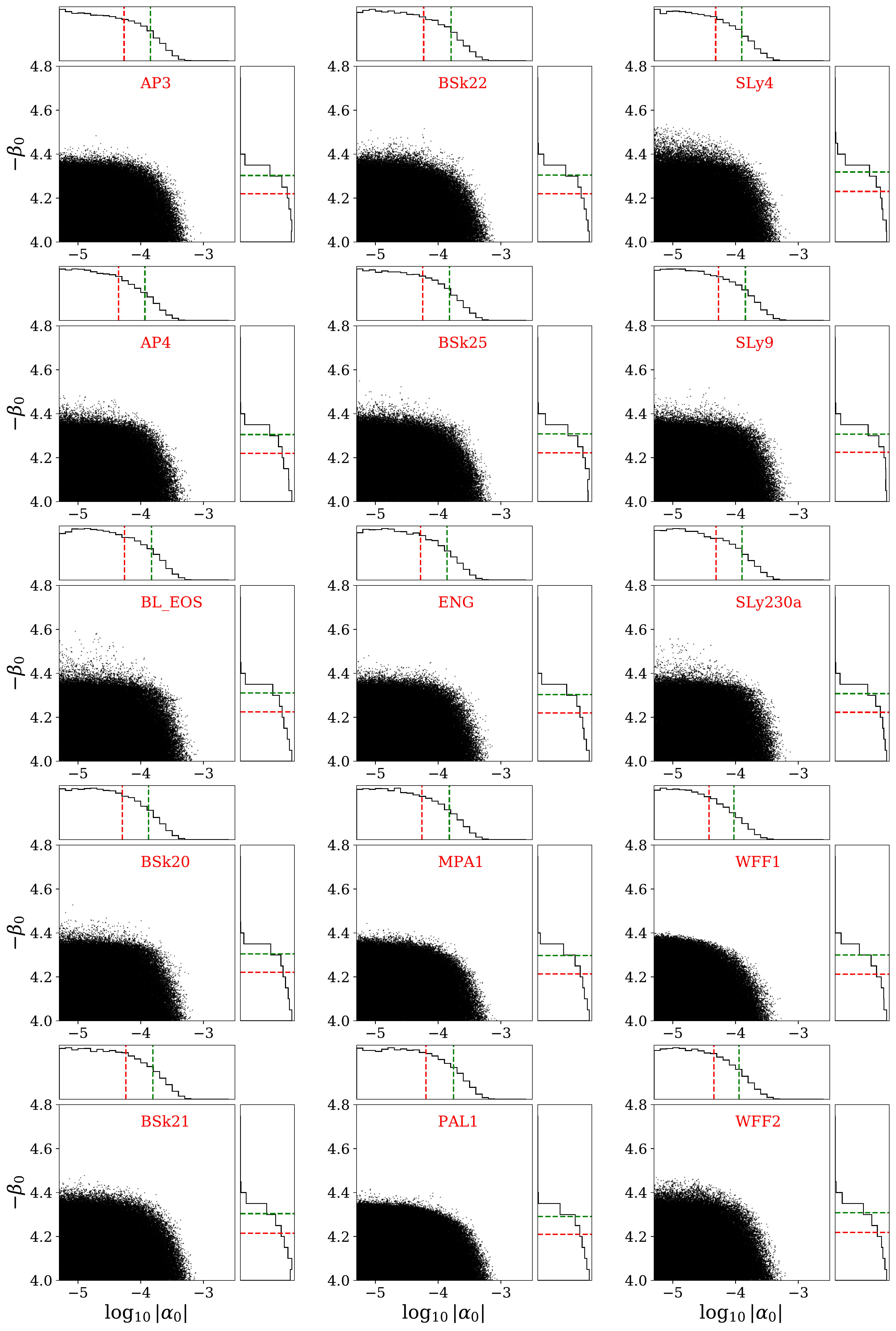}
    \caption{\MG{The marginalized 2-d distribution in the parameter space of
    $\left(\log_{10}|\alpha_0|,-\beta_0\right)$ from MCMC runs on the eight
    pulsars listed in \tabs\ref{tab:psrdat1} and \ref{tab:psrdat2}}{Similar to \fig\ref{fig:mcmc}, but} for all the
    15 EOSs. The marginalized 1-d distributions and the upper limits at 68\% CL
    and 90\% CL are illustrated in upper and right subpanels of each panel.}
    \label{fig:mcmc_all}
\end{figure*}
\begin{figure*}
    \centering
    \includegraphics[width=0.9\linewidth]{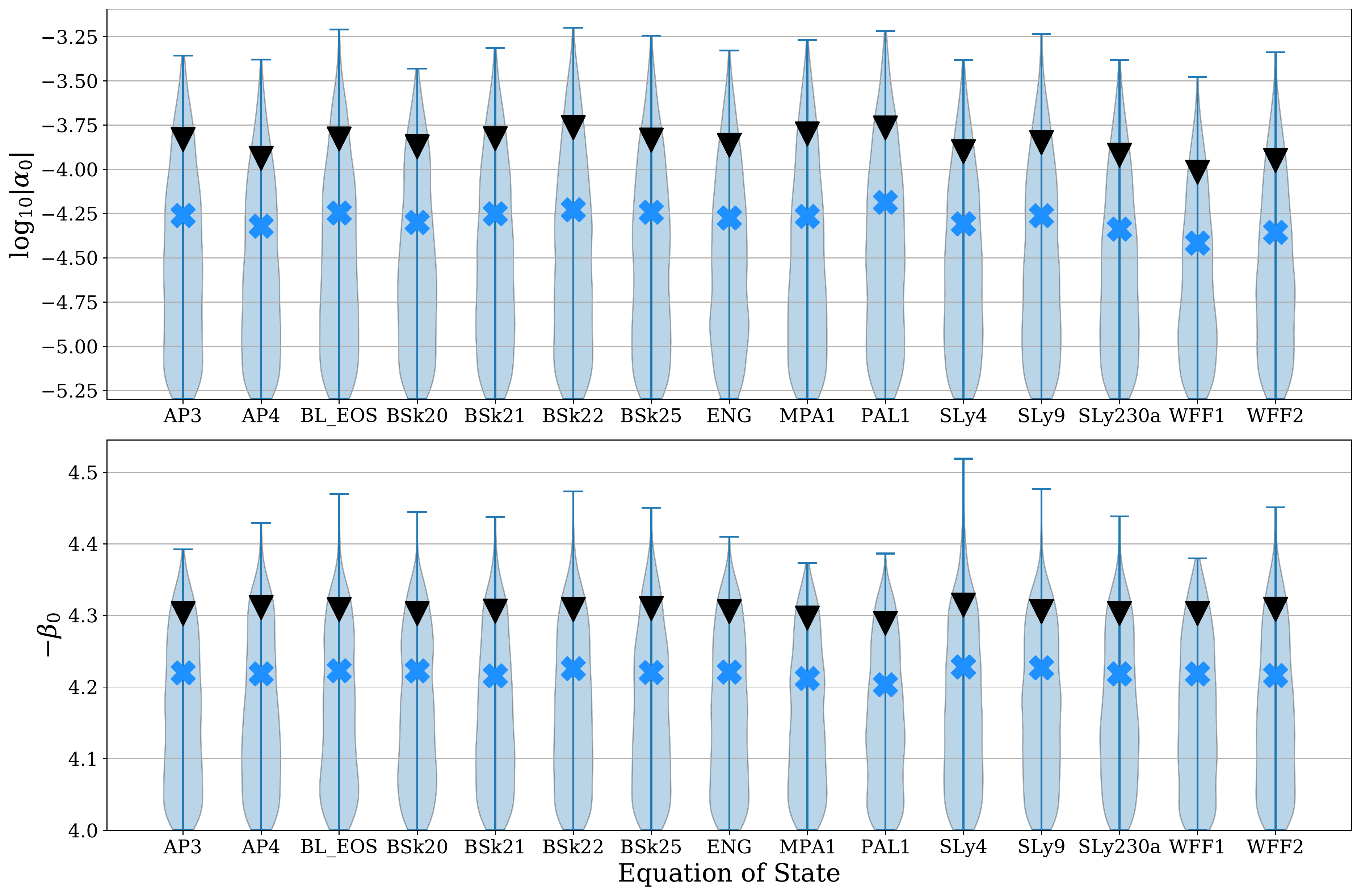}
    \caption{Violin plots of the marginalized 1-d posterior distributions in the
    parameter space of $\log_{10}|\alpha_0|$ (upper panel) and $-\beta_0$ (lower
    panel) for 15 EOSs in our study. The 90\% and 68\% CL upper bounds are shown
    by the black and blue markers, respectively.  }
    \label{fig:mcmc_eos}
\end{figure*}
\begin{figure*}
    \centering
    \includegraphics[width=0.9\linewidth]{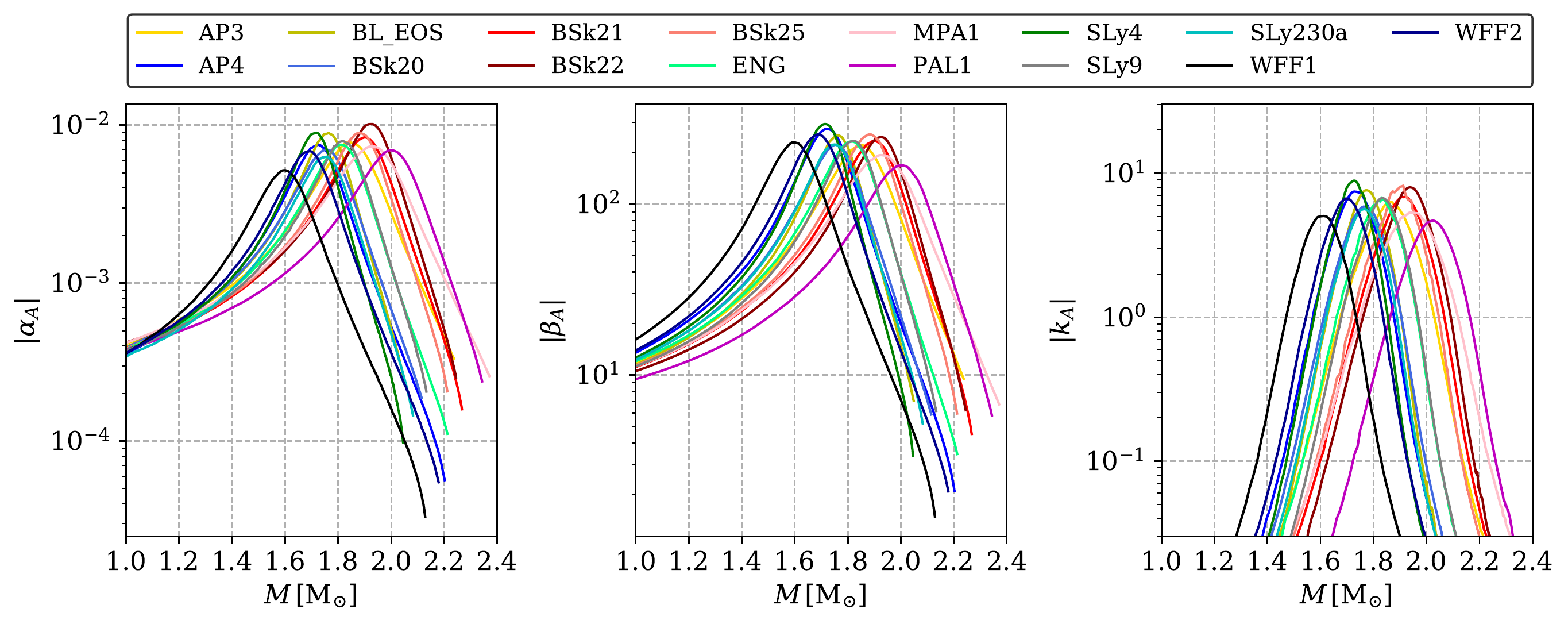}
    \caption{The upper bounds at 90\% CL on the NS effective scalar coupling
    parameters, $\alpha_A$, $\beta_A$, and $k_A$, considering the combined
    constraints from the five NS-WD pulsars in \tab\ref{tab:psrdat1} and three
    NS-NS pulsars in \tab\ref{tab:psrdat2}.}
    \label{fig:90CL}
\end{figure*}

\subsection{Constraints from combining binary pulsars}

\MG{In this subsection, we}{We here} show the constraints from the timing measurements of
binary pulsars from the application of our ROMs. To achieve the speedups of MCMC
calculations, we restrict the priors of $\log_{10}|\alpha_0|$ and $\beta_0$ to
the same ranges as in our ROMs.  With the MCMC simulations done, we can obtain
the posterior distributions of the DEF parameters $\log_{10}|\alpha_0|$ and
$\beta_0$ for the 15 EOSs we adopt.

In \fig\ref{fig:mcmc}, as an example, we show the marginalized 2-dimensional
(2-d) and 1-d posterior distributions in the parameter space of
$\left(\log_{10}|\alpha_0|,\, -\beta_0\right)$ and the constraints at 68\% and
90\% \MG{CLs}{confidence levels (CLs)} for EOS AP4. As mentioned above, the priors of
$\left(\log_{10}|\alpha_0|,\,-\beta_0\right)$ are uniform distributions in the
rectangle region of \fig\ref{fig:mcmc}. The results from pulsars provide very
tight constraints on the parameters of the DEF theory.  The upper limits of
$\log_{10}|\alpha_0|$ and $-\beta_0$ are constrained to $\sim-3.9$ and $\sim4.3$
at 90\% CL, respectively. The constraints are overall similar to previous
results from Refs.~\cite{Zhao2019PhysRevD.100.064034, Shao2017PhRvX...7d1025S}.
The extra double NS systems provide certain but not remarkable improvements in
our results. The tight constraints are mainly contributed by the dipolar
radiation in the NS-WD systems. Here, we mainly aim to provide a test that
proves the ability of our method to constrain these parameters. In the future,
with the further precise observations of binary pulsars, the PPK parameters
$\dot{\omega}$ and $\gamma$ might provide additional constraints for the DEF
gravity. 

We apply the analysis above to all the 15 EOSs.  The marginalized 2-d and 1-d
distributions from the MCMC results for all the 15 EOSs we include in this study
are shown in \fig\ref{fig:mcmc_all}.  In addition, in \fig\ref{fig:mcmc_eos} we
show the 1-d distributions of $\log_{10}|\alpha_0|$ and $-\beta_0$, and their
upper bounds at 68\% and 90\% CLs for all the EOSs. These results are overall
similar, but show the dependence on the choice of EOS. The upper bounds on
$\alpha_0$ at 90\% CL are approximately $10^{-4}$. The upper bounds on
$-\beta_0$ at 90\% CL are $\sim 4.3$. 

Using our ROMs, we can conveniently discuss the spontaneous scalarization
phenomena of a NS.  In \fig\ref{fig:90CL}, we illustrate the upper bounds at
90\% CL on the absolute values of coupling parameters: effective scalar coupling
$\alpha_A$ and its derivative, $\beta_A$, as well as coupling factor for the
moment of inertial, $k_A$, as functions of the NS mass, $m_A$.  The ranges below
the curves mark the regions that are unconstrained up to now.  It is shown
clearly that, for all the 15 EOSs, the allowed maximum values of these
parameters are constrained tightly.  In left panel of \fig\ref{fig:90CL}, the
upper bounds at 90\% CL on the NSs show that for some EOSs, the maximal
effective scalar coupling allowed is $\sim 10^{-2}$. A remarkable scalarization
of NS at this level is still permitted for  a NS with a suitable mass within the
range of $1.6$--$2.0\,\mathrm{M_\odot}$. Compared with earlier
results~\cite{Shao2017PhRvX...7d1025S, Zhao2019PhysRevD.100.064034}, the
``scalarization window'' is still open though slightly limited further.  An
improvement in this study is that, besides the scalar coupling $\alpha_A$, we
can investigate the coupling parameters $\beta_A$ and $k_A$ conveniently with
our ROMs.  As shown in the middle and right panels of \fig\ref{fig:90CL}, the
peaks with extremely large values of $\beta_A$ and $k_A$ shown in
\fig\ref{fig:curve} due to the fact of strong scalarization are essentially
ruled out. Only mild deviation is permitted now. The maximum values are
$\sim3\times10^2$ for $\left|\beta_A\right|$ and $\sim10$ for
$\left|k_A\right|$. Therefore, a very large deviation of these scalar coupling
parameters from GR is not expected based on the contemporary observation.
However, there is still some room for a noteworthy spontaneous scalarization.

\MG{Combining the results,}{In addition,} we note that for a stiffer EOS, spontaneous scalarization
tends to be significant for larger mass, from $1.4$--$1.7\,\mathrm{M_\odot}$
(EOS WFF1) to $1.8$--$2.2\,\mathrm{M_\odot}$ (EOS PAL1).  Most of the NS masses
in our study lie in $m_A \sim 1.3$--$1.7\,\mathrm{M_\odot}$.  As a consequence,
the MCMC simulation shows that, the tightest constraint is obtained for EOS
WFF1.

\MG{In a short summary,}{To summarize,} we provide a quick and statistically reliable method to
constrain the DEF parameters. We verify the validity of the method and give the 
stringent constraints on the DEF theory. The bounds are slightly improved, but
overall similar to previous
studies~\cite{Zhao2019PhysRevD.100.064034,Shao2017PhRvX...7d1025S}. Constraints
on $|\alpha_0|$ and $\beta_0$ depend on the precision of the observation, choice
of pulsar systems, EOSs, and priors.  The ``scalarization window'' is still open
at the level of $\sim10^{-2}$ for all the EOSs.  More observations and suitable
systems in the future, such as possible observations of pulsar timing and GWs
from BNSs with NS masses lying in the ``scalarization window'', will help to
place new constraints on the DEF theory.

\section{\label{sec:conclusion}Conclusion}

\MG{In this paper, we investigate}{This work studies} the DEF theory that predicts large deviations from
GR for NSs through nonperturbative spontaneous scalarization phenomena.  We
briefly review the scalarization phenomena of the DEF theory and relevant
predictions.  To avoid solving the mTOV equations for slowly rotating NSs
repeatedly, which is time-consuming and computationally expensive, we construct
efficient ROMs to derive NSs' properties including $m_A$, $R$, $I_A$,
$\alpha_A$, $\beta_A$, and $k_A$ for the DEF theory\MG{}{, extending earlier work}.  Based on the contemporary
observations about the properties of NSs, such as masses and radii, we build
ROMs for 15 carefully selected EOSs\MG{, extending earlier work}{}.  The ROMs for our
calculations are coded in a python package \pystgromx\ and made public for the
community use. After testing the performance of the ROMs, we find that, compared
with the shooting algorithm, our ROMs can speed up the relevant calculations by
two orders of magnitude for $\alpha_A$ and even three orders for $\beta_A$ and
$k_A$, and yet still keep accuracy at $\sim 1\%$ level. This \MG{}{extended model} gives an accurate
and more comprehensive description of NSs' properties in \MG{}{the} DEF gravity.  

As an application, we utilize \pystgromx\ to explore constraints on the DEF
theory with well-timed binary pulsars through MCMC simulations. We use the
latest precise data from five NS-WD systems and three NS-NS systems, to derive 
tight constraint of the DEF theory. We \MG{bound}{constrain} the DEF parameters to be
$|\alpha_0|\lesssim10^{-3.9}$ and $-\beta_0\lesssim4.3$ at the 90\% CL for a
variety of EOSs. The constraints are dependent on the precision of the
observation, choice of pulsar systems, priors, and EOSs. We find that the
``scalarization window'' still exists at $\sim 10^{-2}$ level. In the future, if
NSs whose masses lie around the ``scalarization window'' can be observed
precisely, the possibility of a strong scalarization for NSs may be excluded
entirely.

In the future, \MG{}{we note that} the new large radio telescopes, e.g., the Five-hundred-meter
Aperture Spherical radio Telescope (FAST)\MG{ in China}{}~\cite{Nan:2011um}  and the
Square Kilometre Array (SKA)\MG{ in Australia and South
Africa}{}~\cite{Kramer:2004hd,Shao:2014wja}, are expected to improve the precision
of pulsar timing and also \MG{}{to} discover new pulsar systems\MG{}{, especially NS-NS systems}. These would help to
close the ``scalarization window''.  Furthermore, the current and
next-generation GW detectors, such as the aLIGO, CE, ET, and KAGRA, could have
more detection of compact binary coalescences in the future. These observations
would provide more information to be used to investigate alternative theories of
gravity in a more precise and accurate way.  Our ROMs are constructed \MG{to satisfy
the requirements of}{for} quick \MG{}{and sound} evaluation of the parameters in the DEF theory,
suitable for applications for binary pulsar experiments, as well as GWs to some
extent.

\acknowledgments 

We are grateful to Norbert Wex for helpful discussions.  This work was
supported by the National SKA Program of China (2020SKA0120300), the National
Natural Science Foundation of China (11975027, 11991053, 11721303), the Young
Elite Scientists Sponsorship Program by the China Association for Science and
Technology (2018QNRC001), the Max Planck Partner Group Program funded by the
Max Planck Society, and the High-performance Computing Platform of Peking
University.

\bibliography{refs}

\providecommand{\noopsort}[1]{}\providecommand{\singleletter}[1]{#1}%
\begin{thebibliography}{78}%
\makeatletter
\providecommand \@ifxundefined [1]{%
 \@ifx{#1\undefined}
}%
\providecommand \@ifnum [1]{%
 \ifnum #1\expandafter \@firstoftwo
 \else \expandafter \@secondoftwo
 \fi
}%
\providecommand \@ifx [1]{%
 \ifx #1\expandafter \@firstoftwo
 \else \expandafter \@secondoftwo
 \fi
}%
\providecommand \natexlab [1]{#1}%
\providecommand \enquote  [1]{``#1''}%
\providecommand \bibnamefont  [1]{#1}%
\providecommand \bibfnamefont [1]{#1}%
\providecommand \citenamefont [1]{#1}%
\providecommand \href@noop [0]{\@secondoftwo}%
\providecommand \href [0]{\begingroup \@sanitize@url \@href}%
\providecommand \@href[1]{\@@startlink{#1}\@@href}%
\providecommand \@@href[1]{\endgroup#1\@@endlink}%
\providecommand \@sanitize@url [0]{\catcode `\\12\catcode `\$12\catcode
  `\&12\catcode `\#12\catcode `\^12\catcode `\_12\catcode `\%12\relax}%
\providecommand \@@startlink[1]{}%
\providecommand \@@endlink[0]{}%
\providecommand \url  [0]{\begingroup\@sanitize@url \@url }%
\providecommand \@url [1]{\endgroup\@href {#1}{\urlprefix }}%
\providecommand \urlprefix  [0]{URL }%
\providecommand \Eprint [0]{\href }%
\providecommand \doibase [0]{http://dx.doi.org/}%
\providecommand \selectlanguage [0]{\@gobble}%
\providecommand \bibinfo  [0]{\@secondoftwo}%
\providecommand \bibfield  [0]{\@secondoftwo}%
\providecommand \translation [1]{[#1]}%
\providecommand \BibitemOpen [0]{}%
\providecommand \bibitemStop [0]{}%
\providecommand \bibitemNoStop [0]{.\EOS\space}%
\providecommand \EOS [0]{\spacefactor3000\relax}%
\providecommand \BibitemShut  [1]{\csname bibitem#1\endcsname}%
\let\auto@bib@innerbib\@empty
\bibitem [{\citenamefont {Einstein}(1915)}]{Einstein:1915ca}%
  \BibitemOpen
  \bibfield  {author} {\bibinfo {author} {\bibfnamefont {A.}~\bibnamefont
  {Einstein}},\ }\href@noop {} {\bibfield  {journal} {\bibinfo  {journal}
  {Sitzungsber. Preuss. Akad. Wiss. Berlin (Math. Phys.)}\ }\textbf {\bibinfo
  {volume} {1915}},\ \bibinfo {pages} {844} (\bibinfo {year}
  {1915})}\BibitemShut {NoStop}%
\bibitem [{\citenamefont {{Will}}(2014)}]{Will2014LRR....17....4W}%
  \BibitemOpen
  \bibfield  {author} {\bibinfo {author} {\bibfnamefont {C.~M.}\ \bibnamefont
  {{Will}}},\ }\href {\doibase 10.12942/lrr-2014-4} {\bibfield  {journal}
  {\bibinfo  {journal} {Living Reviews in Relativity}\ }\textbf {\bibinfo
  {volume} {17}},\ \bibinfo {eid} {4} (\bibinfo {year} {2014})},\ \Eprint
  {http://arxiv.org/abs/1403.7377} {arXiv:1403.7377 [gr-qc]} \BibitemShut
  {NoStop}%
\bibitem [{\citenamefont {Clifton}\ \emph {et~al.}(2012)\citenamefont
  {Clifton}, \citenamefont {Ferreira}, \citenamefont {Padilla},\ and\
  \citenamefont {Skordis}}]{Clifton:2011jh}%
  \BibitemOpen
  \bibfield  {author} {\bibinfo {author} {\bibfnamefont {T.}~\bibnamefont
  {Clifton}}, \bibinfo {author} {\bibfnamefont {P.~G.}\ \bibnamefont
  {Ferreira}}, \bibinfo {author} {\bibfnamefont {A.}~\bibnamefont {Padilla}}, \
  and\ \bibinfo {author} {\bibfnamefont {C.}~\bibnamefont {Skordis}},\ }\href
  {\doibase 10.1016/j.physrep.2012.01.001} {\bibfield  {journal} {\bibinfo
  {journal} {Phys. Rept.}\ }\textbf {\bibinfo {volume} {513}},\ \bibinfo
  {pages} {1} (\bibinfo {year} {2012})},\ \Eprint
  {http://arxiv.org/abs/1106.2476} {arXiv:1106.2476 [astro-ph.CO]} \BibitemShut
  {NoStop}%
\bibitem [{\citenamefont {{Stairs}}(2003)}]{Stairs2003LRR.....6....5S}%
  \BibitemOpen
  \bibfield  {author} {\bibinfo {author} {\bibfnamefont {I.~H.}\ \bibnamefont
  {{Stairs}}},\ }\href {\doibase 10.12942/lrr-2003-5} {\bibfield  {journal}
  {\bibinfo  {journal} {Living Reviews in Relativity}\ }\textbf {\bibinfo
  {volume} {6}},\ \bibinfo {eid} {5} (\bibinfo {year} {2003})},\ \Eprint
  {http://arxiv.org/abs/astro-ph/0307536} {arXiv:astro-ph/0307536 [astro-ph]}
  \BibitemShut {NoStop}%
\bibitem [{\citenamefont {Wex}(2014)}]{Wex:2014nva}%
  \BibitemOpen
  \bibfield  {author} {\bibinfo {author} {\bibfnamefont {N.}~\bibnamefont
  {Wex}},\ }in\ \href@noop {} {\emph {\bibinfo {booktitle} {{Frontiers in
  Relativistic Celestial Mechanics: Applications and Experiments}}}},\
  Vol.~\bibinfo {volume} {2},\ \bibinfo {editor} {edited by\ \bibinfo {editor}
  {\bibfnamefont {S.~M.}\ \bibnamefont {Kopeikin}}}\ (\bibinfo  {publisher}
  {Walter de Gruyter GmbH, Berlin, Boston},\ \bibinfo {year} {2014})\
  p.~\bibinfo {pages} {35},\ \Eprint {http://arxiv.org/abs/1402.5594}
  {arXiv:1402.5594 [gr-qc]} \BibitemShut {NoStop}%
\bibitem [{\citenamefont {Shao}\ and\ \citenamefont
  {Wex}(2016)}]{Shao:2016ezh}%
  \BibitemOpen
  \bibfield  {author} {\bibinfo {author} {\bibfnamefont {L.}~\bibnamefont
  {Shao}}\ and\ \bibinfo {author} {\bibfnamefont {N.}~\bibnamefont {Wex}},\
  }\href {\doibase 10.1007/s11433-016-0087-6} {\bibfield  {journal} {\bibinfo
  {journal} {Sci. China Phys. Mech. Astron.}\ }\textbf {\bibinfo {volume}
  {59}},\ \bibinfo {pages} {699501} (\bibinfo {year} {2016})},\ \Eprint
  {http://arxiv.org/abs/1604.03662} {arXiv:1604.03662 [gr-qc]} \BibitemShut
  {NoStop}%
\bibitem [{\citenamefont {Abbott}\ \emph {et~al.}(2016)\citenamefont {Abbott}
  \emph {et~al.}}]{TheLIGOScientific:2016src}%
  \BibitemOpen
  \bibfield  {author} {\bibinfo {author} {\bibfnamefont {B.}~\bibnamefont
  {Abbott}} \emph {et~al.} (\bibinfo {collaboration} {LIGO Scientific and Virgo
  Collaborations}),\ }\href {\doibase 10.1103/PhysRevLett.116.221101}
  {\bibfield  {journal} {\bibinfo  {journal} {Phys. Rev. Lett.}\ }\textbf
  {\bibinfo {volume} {116}},\ \bibinfo {pages} {221101} (\bibinfo {year}
  {2016})},\ \bibinfo {note} {[Erratum: Phys.Rev.Lett. 121, 129902 (2018)]},\
  \Eprint {http://arxiv.org/abs/1602.03841} {arXiv:1602.03841 [gr-qc]}
  \BibitemShut {NoStop}%
\bibitem [{\citenamefont {Abbott}\ \emph
  {et~al.}(2017{\natexlab{a}})\citenamefont {Abbott} \emph
  {et~al.}}]{Abbott:2017vtc}%
  \BibitemOpen
  \bibfield  {author} {\bibinfo {author} {\bibfnamefont {B.~P.}\ \bibnamefont
  {Abbott}} \emph {et~al.} (\bibinfo {collaboration} {LIGO Scientific and Virgo
  Collaborations}),\ }\href {\doibase 10.1103/PhysRevLett.118.221101}
  {\bibfield  {journal} {\bibinfo  {journal} {Phys. Rev. Lett.}\ }\textbf
  {\bibinfo {volume} {118}},\ \bibinfo {pages} {221101} (\bibinfo {year}
  {2017}{\natexlab{a}})},\ \bibinfo {note} {[Erratum: Phys.Rev.Lett. 121,
  129901 (2018)]},\ \Eprint {http://arxiv.org/abs/1706.01812} {arXiv:1706.01812
  [gr-qc]} \BibitemShut {NoStop}%
\bibitem [{\citenamefont {Abbott}\ \emph
  {et~al.}(2019{\natexlab{a}})\citenamefont {Abbott} \emph
  {et~al.}}]{LIGOScientific:2019fpa}%
  \BibitemOpen
  \bibfield  {author} {\bibinfo {author} {\bibfnamefont {B.}~\bibnamefont
  {Abbott}} \emph {et~al.} (\bibinfo {collaboration} {LIGO Scientific and Virgo
  Collaborations}),\ }\href {\doibase 10.1103/PhysRevD.100.104036} {\bibfield
  {journal} {\bibinfo  {journal} {Phys. Rev. D}\ }\textbf {\bibinfo {volume}
  {100}},\ \bibinfo {pages} {104036} (\bibinfo {year} {2019}{\natexlab{a}})},\
  \Eprint {http://arxiv.org/abs/1903.04467} {arXiv:1903.04467 [gr-qc]}
  \BibitemShut {NoStop}%
\bibitem [{\citenamefont {Abbott}\ \emph
  {et~al.}(2021{\natexlab{a}})\citenamefont {Abbott} \emph
  {et~al.}}]{LIGOScientific:2020tif}%
  \BibitemOpen
  \bibfield  {author} {\bibinfo {author} {\bibfnamefont {R.}~\bibnamefont
  {Abbott}} \emph {et~al.} (\bibinfo {collaboration} {LIGO Scientific and Virgo
  Collaborations}),\ }\href {\doibase 10.1103/PhysRevD.103.122002} {\bibfield
  {journal} {\bibinfo  {journal} {Phys. Rev. D}\ }\textbf {\bibinfo {volume}
  {103}},\ \bibinfo {pages} {122002} (\bibinfo {year} {2021}{\natexlab{a}})},\
  \Eprint {http://arxiv.org/abs/2010.14529} {arXiv:2010.14529 [gr-qc]}
  \BibitemShut {NoStop}%
\bibitem [{\citenamefont {Abbott}\ \emph
  {et~al.}(2017{\natexlab{b}})\citenamefont {Abbott} \emph
  {et~al.}}]{TheLIGOScientific:2017qsa}%
  \BibitemOpen
  \bibfield  {author} {\bibinfo {author} {\bibfnamefont {B.}~\bibnamefont
  {Abbott}} \emph {et~al.} (\bibinfo {collaboration} {LIGO Scientific and Virgo
  Collaborations}),\ }\href {\doibase 10.1103/PhysRevLett.119.161101}
  {\bibfield  {journal} {\bibinfo  {journal} {Phys. Rev. Lett.}\ }\textbf
  {\bibinfo {volume} {119}},\ \bibinfo {pages} {161101} (\bibinfo {year}
  {2017}{\natexlab{b}})},\ \Eprint {http://arxiv.org/abs/1710.05832}
  {arXiv:1710.05832 [gr-qc]} \BibitemShut {NoStop}%
\bibitem [{\citenamefont {Abbott}\ \emph
  {et~al.}(2017{\natexlab{c}})\citenamefont {Abbott} \emph
  {et~al.}}]{GBM:2017lvd}%
  \BibitemOpen
  \bibfield  {author} {\bibinfo {author} {\bibfnamefont {B.~P.}\ \bibnamefont
  {Abbott}} \emph {et~al.},\ }\href {\doibase 10.3847/2041-8213/aa91c9}
  {\bibfield  {journal} {\bibinfo  {journal} {Astrophys. J. Lett.}\ }\textbf
  {\bibinfo {volume} {848}},\ \bibinfo {pages} {L12} (\bibinfo {year}
  {2017}{\natexlab{c}})},\ \Eprint {http://arxiv.org/abs/1710.05833}
  {arXiv:1710.05833 [astro-ph.HE]} \BibitemShut {NoStop}%
\bibitem [{\citenamefont {Abbott}\ \emph
  {et~al.}(2019{\natexlab{b}})\citenamefont {Abbott} \emph
  {et~al.}}]{Abbott:2018lct}%
  \BibitemOpen
  \bibfield  {author} {\bibinfo {author} {\bibfnamefont {B.}~\bibnamefont
  {Abbott}} \emph {et~al.} (\bibinfo {collaboration} {LIGO Scientific and Virgo
  Collaborations}),\ }\href {\doibase 10.1103/PhysRevLett.123.011102}
  {\bibfield  {journal} {\bibinfo  {journal} {Phys. Rev. Lett.}\ }\textbf
  {\bibinfo {volume} {123}},\ \bibinfo {pages} {011102} (\bibinfo {year}
  {2019}{\natexlab{b}})},\ \Eprint {http://arxiv.org/abs/1811.00364}
  {arXiv:1811.00364 [gr-qc]} \BibitemShut {NoStop}%
\bibitem [{\citenamefont {Berti}\ \emph {et~al.}(2015)\citenamefont {Berti}
  \emph {et~al.}}]{Berti:2015itd}%
  \BibitemOpen
  \bibfield  {author} {\bibinfo {author} {\bibfnamefont {E.}~\bibnamefont
  {Berti}} \emph {et~al.},\ }\href {\doibase 10.1088/0264-9381/32/24/243001}
  {\bibfield  {journal} {\bibinfo  {journal} {Class. Quant. Grav.}\ }\textbf
  {\bibinfo {volume} {32}},\ \bibinfo {pages} {243001} (\bibinfo {year}
  {2015})},\ \Eprint {http://arxiv.org/abs/1501.07274} {arXiv:1501.07274
  [gr-qc]} \BibitemShut {NoStop}%
\bibitem [{\citenamefont {Will}(2018)}]{Will:2018bme}%
  \BibitemOpen
  \bibfield  {author} {\bibinfo {author} {\bibfnamefont {C.~M.}\ \bibnamefont
  {Will}},\ }\href
  {https://www.cambridge.org/academic/subjects/physics/cosmology-relativity-and-gravitation/theory-and-experiment-gravitational-physics-2nd-edition?format=AR&isbn=9781108679824}
  {\emph {\bibinfo {title} {{Theory and Experiment in Gravitational
  Physics}}}}\ (\bibinfo  {publisher} {Cambridge University Press},\ \bibinfo
  {address} {Cambridge, England},\ \bibinfo {year} {2018})\BibitemShut
  {NoStop}%
\bibitem [{\citenamefont {Kaluza}(2018)}]{Kaluza:1921tu}%
  \BibitemOpen
  \bibfield  {author} {\bibinfo {author} {\bibfnamefont {T.}~\bibnamefont
  {Kaluza}},\ }\href {\doibase 10.1142/S0218271818700017} {\bibfield  {journal}
  {\bibinfo  {journal} {Int. J. Mod. Phys. D}\ }\textbf {\bibinfo {volume}
  {27}},\ \bibinfo {pages} {1870001} (\bibinfo {year} {2018})},\ \Eprint
  {http://arxiv.org/abs/1803.08616} {arXiv:1803.08616 [physics.hist-ph]}
  \BibitemShut {NoStop}%
\bibitem [{\citenamefont {{Klein}}(1926)}]{Klein1926ZPhy...37..895K}%
  \BibitemOpen
  \bibfield  {author} {\bibinfo {author} {\bibfnamefont {O.}~\bibnamefont
  {{Klein}}},\ }\href {\doibase 10.1007/BF01397481} {\bibfield  {journal}
  {\bibinfo  {journal} {Zeitschrift fur Physik}\ }\textbf {\bibinfo {volume}
  {37}},\ \bibinfo {pages} {895} (\bibinfo {year} {1926})}\BibitemShut
  {NoStop}%
\bibitem [{\citenamefont {Fujii}\ and\ \citenamefont
  {Maeda}(2007)}]{Fujii:2003pa}%
  \BibitemOpen
  \bibfield  {author} {\bibinfo {author} {\bibfnamefont {Y.}~\bibnamefont
  {Fujii}}\ and\ \bibinfo {author} {\bibfnamefont {K.}~\bibnamefont {Maeda}},\
  }\href {\doibase 10.1017/CBO9780511535093} {\emph {\bibinfo {title} {{The
  Scalar-tensor Theory of Gravitation}}}},\ Cambridge Monographs on
  Mathematical Physics\ (\bibinfo  {publisher} {Cambridge University Press,
  Cambridge, England},\ \bibinfo {year} {2007})\BibitemShut {NoStop}%
\bibitem [{\citenamefont {Goenner}(2012)}]{Goenner:2012cq}%
  \BibitemOpen
  \bibfield  {author} {\bibinfo {author} {\bibfnamefont {H.}~\bibnamefont
  {Goenner}},\ }\href {\doibase 10.1007/s10714-012-1378-8} {\bibfield
  {journal} {\bibinfo  {journal} {Gen. Rel. Grav.}\ }\textbf {\bibinfo {volume}
  {44}},\ \bibinfo {pages} {2077} (\bibinfo {year} {2012})},\ \Eprint
  {http://arxiv.org/abs/1204.3455} {arXiv:1204.3455 [gr-qc]} \BibitemShut
  {NoStop}%
\bibitem [{\citenamefont {Jordan}(1949)}]{Jordan:1949zz}%
  \BibitemOpen
  \bibfield  {author} {\bibinfo {author} {\bibfnamefont {P.}~\bibnamefont
  {Jordan}},\ }\href {\doibase 10.1038/164637a0} {\bibfield  {journal}
  {\bibinfo  {journal} {Nature}\ }\textbf {\bibinfo {volume} {164}},\ \bibinfo
  {pages} {637} (\bibinfo {year} {1949})}\BibitemShut {NoStop}%
\bibitem [{\citenamefont {Jordan}(1959)}]{Jordan:1959eg}%
  \BibitemOpen
  \bibfield  {author} {\bibinfo {author} {\bibfnamefont {P.}~\bibnamefont
  {Jordan}},\ }\href {\doibase 10.1007/BF01375155} {\bibfield  {journal}
  {\bibinfo  {journal} {Z. Phys.}\ }\textbf {\bibinfo {volume} {157}},\
  \bibinfo {pages} {112} (\bibinfo {year} {1959})}\BibitemShut {NoStop}%
\bibitem [{\citenamefont {Fierz}(1956)}]{Fierz:1956zz}%
  \BibitemOpen
  \bibfield  {author} {\bibinfo {author} {\bibfnamefont {M.}~\bibnamefont
  {Fierz}},\ }\href@noop {} {\bibfield  {journal} {\bibinfo  {journal} {Helv.
  Phys. Acta}\ }\textbf {\bibinfo {volume} {29}},\ \bibinfo {pages} {128}
  (\bibinfo {year} {1956})}\BibitemShut {NoStop}%
\bibitem [{\citenamefont {Brans}\ and\ \citenamefont
  {Dicke}(1961)}]{Brans:1961sx}%
  \BibitemOpen
  \bibfield  {author} {\bibinfo {author} {\bibfnamefont {C.}~\bibnamefont
  {Brans}}\ and\ \bibinfo {author} {\bibfnamefont {R.}~\bibnamefont {Dicke}},\
  }\href {\doibase 10.1103/PhysRev.124.925} {\bibfield  {journal} {\bibinfo
  {journal} {Phys. Rev.}\ }\textbf {\bibinfo {volume} {124}},\ \bibinfo {pages}
  {925} (\bibinfo {year} {1961})}\BibitemShut {NoStop}%
\bibitem [{\citenamefont {{Damour}}\ and\ \citenamefont
  {{Esposito-Far\`ese}}(1992)}]{Damour1992CQGra...9.2093D}%
  \BibitemOpen
  \bibfield  {author} {\bibinfo {author} {\bibfnamefont {T.}~\bibnamefont
  {{Damour}}}\ and\ \bibinfo {author} {\bibfnamefont {G.}~\bibnamefont
  {{Esposito-Far\`ese}}},\ }\href {\doibase 10.1088/0264-9381/9/9/015}
  {\bibfield  {journal} {\bibinfo  {journal} {Classical and Quantum Gravity}\
  }\textbf {\bibinfo {volume} {9}},\ \bibinfo {pages} {2093} (\bibinfo {year}
  {1992})}\BibitemShut {NoStop}%
\bibitem [{\citenamefont {Damour}\ and\ \citenamefont
  {Esposito-Far\`ese}(1993)}]{Damour1993PhysRevLett.70.2220}%
  \BibitemOpen
  \bibfield  {author} {\bibinfo {author} {\bibfnamefont {T.}~\bibnamefont
  {Damour}}\ and\ \bibinfo {author} {\bibfnamefont {G.}~\bibnamefont
  {Esposito-Far\`ese}},\ }\href {\doibase 10.1103/PhysRevLett.70.2220}
  {\bibfield  {journal} {\bibinfo  {journal} {Phys. Rev. Lett.}\ }\textbf
  {\bibinfo {volume} {70}},\ \bibinfo {pages} {2220} (\bibinfo {year}
  {1993})}\BibitemShut {NoStop}%
\bibitem [{\citenamefont {Damour}\ and\ \citenamefont
  {Esposito-Far\`ese}(1996)}]{Damour1996PhysRevD.54.1474}%
  \BibitemOpen
  \bibfield  {author} {\bibinfo {author} {\bibfnamefont {T.}~\bibnamefont
  {Damour}}\ and\ \bibinfo {author} {\bibfnamefont {G.}~\bibnamefont
  {Esposito-Far\`ese}},\ }\href {\doibase 10.1103/PhysRevD.54.1474} {\bibfield
  {journal} {\bibinfo  {journal} {Phys. Rev. D}\ }\textbf {\bibinfo {volume}
  {54}},\ \bibinfo {pages} {1474} (\bibinfo {year} {1996})}\BibitemShut
  {NoStop}%
\bibitem [{\citenamefont {Bertotti}\ \emph {et~al.}(2003)\citenamefont
  {Bertotti}, \citenamefont {Iess},\ and\ \citenamefont
  {Tortora}}]{Bertotti:2003rm}%
  \BibitemOpen
  \bibfield  {author} {\bibinfo {author} {\bibfnamefont {B.}~\bibnamefont
  {Bertotti}}, \bibinfo {author} {\bibfnamefont {L.}~\bibnamefont {Iess}}, \
  and\ \bibinfo {author} {\bibfnamefont {P.}~\bibnamefont {Tortora}},\ }\href
  {\doibase 10.1038/nature01997} {\bibfield  {journal} {\bibinfo  {journal}
  {Nature}\ }\textbf {\bibinfo {volume} {425}},\ \bibinfo {pages} {374}
  (\bibinfo {year} {2003})}\BibitemShut {NoStop}%
\bibitem [{\citenamefont {{Shao}}\ \emph {et~al.}(2017)\citenamefont {{Shao}},
  \citenamefont {{Sennett}}, \citenamefont {{Buonanno}}, \citenamefont
  {{Kramer}},\ and\ \citenamefont {{Wex}}}]{Shao2017PhRvX...7d1025S}%
  \BibitemOpen
  \bibfield  {author} {\bibinfo {author} {\bibfnamefont {L.}~\bibnamefont
  {{Shao}}}, \bibinfo {author} {\bibfnamefont {N.}~\bibnamefont {{Sennett}}},
  \bibinfo {author} {\bibfnamefont {A.}~\bibnamefont {{Buonanno}}}, \bibinfo
  {author} {\bibfnamefont {M.}~\bibnamefont {{Kramer}}}, \ and\ \bibinfo
  {author} {\bibfnamefont {N.}~\bibnamefont {{Wex}}},\ }\href {\doibase
  10.1103/PhysRevX.7.041025} {\bibfield  {journal} {\bibinfo  {journal}
  {Physical Review X}\ }\textbf {\bibinfo {volume} {7}},\ \bibinfo {eid}
  {041025} (\bibinfo {year} {2017})},\ \Eprint
  {http://arxiv.org/abs/1704.07561} {arXiv:1704.07561 [gr-qc]} \BibitemShut
  {NoStop}%
\bibitem [{\citenamefont {Sennett}\ \emph {et~al.}(2017)\citenamefont
  {Sennett}, \citenamefont {Shao},\ and\ \citenamefont
  {Steinhoff}}]{Sennett:2017lcx}%
  \BibitemOpen
  \bibfield  {author} {\bibinfo {author} {\bibfnamefont {N.}~\bibnamefont
  {Sennett}}, \bibinfo {author} {\bibfnamefont {L.}~\bibnamefont {Shao}}, \
  and\ \bibinfo {author} {\bibfnamefont {J.}~\bibnamefont {Steinhoff}},\ }\href
  {\doibase 10.1103/PhysRevD.96.084019} {\bibfield  {journal} {\bibinfo
  {journal} {Phys. Rev. D}\ }\textbf {\bibinfo {volume} {96}},\ \bibinfo
  {pages} {084019} (\bibinfo {year} {2017})},\ \Eprint
  {http://arxiv.org/abs/1708.08285} {arXiv:1708.08285 [gr-qc]} \BibitemShut
  {NoStop}%
\bibitem [{\citenamefont {{Damour}}(2009)}]{Damour2009ASSL..359....1D}%
  \BibitemOpen
  \bibfield  {author} {\bibinfo {author} {\bibfnamefont {T.}~\bibnamefont
  {{Damour}}},\ }in\ \href@noop {} {\emph {\bibinfo {booktitle} {{Physics of
  Relativistic Objects in Compact Binaries: From Birth to Coalescence}}}},\
  Vol.\ \bibinfo {volume} {359},\ \bibinfo {editor} {edited by\ \bibinfo
  {editor} {\bibfnamefont {M.}~\bibnamefont {Colpi}}, \bibinfo {editor}
  {\bibfnamefont {P.}~\bibnamefont {Casella}}, \bibinfo {editor} {\bibfnamefont
  {V.}~\bibnamefont {Gorini}}, \bibinfo {editor} {\bibfnamefont
  {U.}~\bibnamefont {Moschella}}, \ and\ \bibinfo {editor} {\bibfnamefont
  {A.}~\bibnamefont {Possenti}}}\ (\bibinfo  {publisher} {Springer,
  Dordrecht},\ \bibinfo {year} {2009})\ p.~\bibinfo {pages} {1},\ \Eprint
  {http://arxiv.org/abs/0704.0749} {arXiv:0704.0749 [gr-qc]} \BibitemShut
  {NoStop}%
\bibitem [{\citenamefont {Anderson}\ \emph {et~al.}(2019)\citenamefont
  {Anderson}, \citenamefont {Freire},\ and\ \citenamefont
  {Yunes}}]{Anderson:2019eay}%
  \BibitemOpen
  \bibfield  {author} {\bibinfo {author} {\bibfnamefont {D.}~\bibnamefont
  {Anderson}}, \bibinfo {author} {\bibfnamefont {P.}~\bibnamefont {Freire}}, \
  and\ \bibinfo {author} {\bibfnamefont {N.}~\bibnamefont {Yunes}},\ }\href
  {\doibase 10.1088/1361-6382/ab3a1c} {\bibfield  {journal} {\bibinfo
  {journal} {Class. Quant. Grav.}\ }\textbf {\bibinfo {volume} {36}},\ \bibinfo
  {pages} {225009} (\bibinfo {year} {2019})},\ \Eprint
  {http://arxiv.org/abs/1901.00938} {arXiv:1901.00938 [gr-qc]} \BibitemShut
  {NoStop}%
\bibitem [{\citenamefont {Shao}(2019{\natexlab{a}})}]{Shao:2019nso}%
  \BibitemOpen
  \bibfield  {author} {\bibinfo {author} {\bibfnamefont {L.}~\bibnamefont
  {Shao}},\ }in\ \href {\doibase 10.1142/9789811213984_0043} {\emph {\bibinfo
  {booktitle} {{8th Meeting on CPT and Lorentz Symmetry}}}}\ (\bibinfo {year}
  {2019})\ \Eprint {http://arxiv.org/abs/1905.08405} {arXiv:1905.08405 [gr-qc]}
  \BibitemShut {NoStop}%
\bibitem [{\citenamefont {Damour}\ and\ \citenamefont
  {Taylor}(1992)}]{Damour1992PhysRevD.45.1840}%
  \BibitemOpen
  \bibfield  {author} {\bibinfo {author} {\bibfnamefont {T.}~\bibnamefont
  {Damour}}\ and\ \bibinfo {author} {\bibfnamefont {J.~H.}\ \bibnamefont
  {Taylor}},\ }\href {\doibase 10.1103/PhysRevD.45.1840} {\bibfield  {journal}
  {\bibinfo  {journal} {Phys. Rev. D}\ }\textbf {\bibinfo {volume} {45}},\
  \bibinfo {pages} {1840} (\bibinfo {year} {1992})}\BibitemShut {NoStop}%
\bibitem [{\citenamefont {Abbott}\ \emph
  {et~al.}(2021{\natexlab{b}})\citenamefont {Abbott} \emph
  {et~al.}}]{LIGOScientific:2020ibl}%
  \BibitemOpen
  \bibfield  {author} {\bibinfo {author} {\bibfnamefont {R.}~\bibnamefont
  {Abbott}} \emph {et~al.} (\bibinfo {collaboration} {LIGO Scientific and Virgo
  Collaborations}),\ }\href {\doibase 10.1103/PhysRevX.11.021053} {\bibfield
  {journal} {\bibinfo  {journal} {Phys. Rev. X}\ }\textbf {\bibinfo {volume}
  {11}},\ \bibinfo {pages} {021053} (\bibinfo {year} {2021}{\natexlab{b}})},\
  \Eprint {http://arxiv.org/abs/2010.14527} {arXiv:2010.14527 [gr-qc]}
  \BibitemShut {NoStop}%
\bibitem [{\citenamefont {Finn}(1992)}]{Finn:1992wt}%
  \BibitemOpen
  \bibfield  {author} {\bibinfo {author} {\bibfnamefont {L.~S.}\ \bibnamefont
  {Finn}},\ }\href {\doibase 10.1103/PhysRevD.46.5236} {\bibfield  {journal}
  {\bibinfo  {journal} {Phys. Rev.}\ }\textbf {\bibinfo {volume} {D46}},\
  \bibinfo {pages} {5236} (\bibinfo {year} {1992})},\ \Eprint
  {http://arxiv.org/abs/gr-qc/9209010} {arXiv:gr-qc/9209010 [gr-qc]}
  \BibitemShut {NoStop}%
\bibitem [{\citenamefont {Zhao}\ \emph {et~al.}(2019)\citenamefont {Zhao},
  \citenamefont {Shao}, \citenamefont {Cao},\ and\ \citenamefont
  {Ma}}]{Zhao2019PhysRevD.100.064034}%
  \BibitemOpen
  \bibfield  {author} {\bibinfo {author} {\bibfnamefont {J.}~\bibnamefont
  {Zhao}}, \bibinfo {author} {\bibfnamefont {L.}~\bibnamefont {Shao}}, \bibinfo
  {author} {\bibfnamefont {Z.}~\bibnamefont {Cao}}, \ and\ \bibinfo {author}
  {\bibfnamefont {B.-Q.}\ \bibnamefont {Ma}},\ }\href {\doibase
  10.1103/PhysRevD.100.064034} {\bibfield  {journal} {\bibinfo  {journal}
  {Phys. Rev. D}\ }\textbf {\bibinfo {volume} {100}},\ \bibinfo {pages}
  {064034} (\bibinfo {year} {2019})}\BibitemShut {NoStop}%
\bibitem [{\citenamefont {Abbott}\ \emph
  {et~al.}(2017{\natexlab{d}})\citenamefont {Abbott} \emph
  {et~al.}}]{Evans:2016mbw-CE}%
  \BibitemOpen
  \bibfield  {author} {\bibinfo {author} {\bibfnamefont {B.~P.}\ \bibnamefont
  {Abbott}} \emph {et~al.} (\bibinfo {collaboration} {LIGO Scientific
  Collaboration}),\ }\href {\doibase 10.1088/1361-6382/aa51f4} {\bibfield
  {journal} {\bibinfo  {journal} {Class. Quant. Grav.}\ }\textbf {\bibinfo
  {volume} {34}},\ \bibinfo {pages} {044001} (\bibinfo {year}
  {2017}{\natexlab{d}})},\ \Eprint {http://arxiv.org/abs/1607.08697}
  {arXiv:1607.08697 [astro-ph.IM]} \BibitemShut {NoStop}%
\bibitem [{\citenamefont {Hild}\ \emph {et~al.}(2011)\citenamefont {Hild} \emph
  {et~al.}}]{Hild:2010id-ET}%
  \BibitemOpen
  \bibfield  {author} {\bibinfo {author} {\bibfnamefont {S.}~\bibnamefont
  {Hild}} \emph {et~al.},\ }\href {\doibase 10.1088/0264-9381/28/9/094013}
  {\bibfield  {journal} {\bibinfo  {journal} {Class. Quant. Grav.}\ }\textbf
  {\bibinfo {volume} {28}},\ \bibinfo {pages} {094013} (\bibinfo {year}
  {2011})},\ \Eprint {http://arxiv.org/abs/1012.0908} {arXiv:1012.0908 [gr-qc]}
  \BibitemShut {NoStop}%
\bibitem [{\citenamefont {Yagi}\ and\ \citenamefont
  {Tanaka}(2010)}]{Yagi:2009zz}%
  \BibitemOpen
  \bibfield  {author} {\bibinfo {author} {\bibfnamefont {K.}~\bibnamefont
  {Yagi}}\ and\ \bibinfo {author} {\bibfnamefont {T.}~\bibnamefont {Tanaka}},\
  }\href {\doibase 10.1143/PTP.123.1069} {\bibfield  {journal} {\bibinfo
  {journal} {Prog. Theor. Phys.}\ }\textbf {\bibinfo {volume} {123}},\ \bibinfo
  {pages} {1069} (\bibinfo {year} {2010})},\ \Eprint
  {http://arxiv.org/abs/0908.3283} {arXiv:0908.3283 [gr-qc]} \BibitemShut
  {NoStop}%
\bibitem [{\citenamefont {Sedda}\ \emph {et~al.}(2020)\citenamefont {Sedda}
  \emph {et~al.}}]{Sedda2020CQGra..37u5011A}%
  \BibitemOpen
  \bibfield  {author} {\bibinfo {author} {\bibfnamefont {M.~A.}\ \bibnamefont
  {Sedda}} \emph {et~al.},\ }\href {\doibase 10.1088/1361-6382/abb5c1}
  {\bibfield  {journal} {\bibinfo  {journal} {Class. Quant. Grav.}\ }\textbf
  {\bibinfo {volume} {37}},\ \bibinfo {pages} {215011} (\bibinfo {year}
  {2020})},\ \Eprint {http://arxiv.org/abs/1908.11375} {arXiv:1908.11375
  [gr-qc]} \BibitemShut {NoStop}%
\bibitem [{\citenamefont {Arca~Sedda}\ \emph {et~al.}(2021)\citenamefont
  {Arca~Sedda} \emph {et~al.}}]{Sedda:2021hpg}%
  \BibitemOpen
  \bibfield  {author} {\bibinfo {author} {\bibfnamefont {M.}~\bibnamefont
  {Arca~Sedda}} \emph {et~al.},\ }\href {\doibase 10.1007/s10686-021-09713-z}
  {\bibfield  {journal} {\bibinfo  {journal} {Exp. Astron.}\ } (\bibinfo {year}
  {2021}),\ 10.1007/s10686-021-09713-z},\ \Eprint
  {http://arxiv.org/abs/2104.14583} {arXiv:2104.14583 [gr-qc]} \BibitemShut
  {NoStop}%
\bibitem [{\citenamefont {Liu}\ \emph {et~al.}(2020{\natexlab{a}})\citenamefont
  {Liu}, \citenamefont {Shao}, \citenamefont {Zhao},\ and\ \citenamefont
  {Gao}}]{Liu:2020nwz}%
  \BibitemOpen
  \bibfield  {author} {\bibinfo {author} {\bibfnamefont {C.}~\bibnamefont
  {Liu}}, \bibinfo {author} {\bibfnamefont {L.}~\bibnamefont {Shao}}, \bibinfo
  {author} {\bibfnamefont {J.}~\bibnamefont {Zhao}}, \ and\ \bibinfo {author}
  {\bibfnamefont {Y.}~\bibnamefont {Gao}},\ }\href {\doibase
  10.1093/mnras/staa1512} {\bibfield  {journal} {\bibinfo  {journal} {Mon. Not.
  Roy. Astron. Soc.}\ }\textbf {\bibinfo {volume} {496}},\ \bibinfo {pages}
  {182} (\bibinfo {year} {2020}{\natexlab{a}})},\ \Eprint
  {http://arxiv.org/abs/2004.12096} {arXiv:2004.12096 [astro-ph.HE]}
  \BibitemShut {NoStop}%
\bibitem [{\citenamefont {Lattimer}\ and\ \citenamefont
  {Prakash}(2001)}]{Lattimer:2000nx}%
  \BibitemOpen
  \bibfield  {author} {\bibinfo {author} {\bibfnamefont {J.~M.}\ \bibnamefont
  {Lattimer}}\ and\ \bibinfo {author} {\bibfnamefont {M.}~\bibnamefont
  {Prakash}},\ }\href {\doibase 10.1086/319702} {\bibfield  {journal} {\bibinfo
   {journal} {Astrophys. J.}\ }\textbf {\bibinfo {volume} {550}},\ \bibinfo
  {pages} {426} (\bibinfo {year} {2001})},\ \Eprint
  {http://arxiv.org/abs/astro-ph/0002232} {arXiv:astro-ph/0002232 [astro-ph]}
  \BibitemShut {NoStop}%
\bibitem [{\citenamefont {Shao}(2019{\natexlab{b}})}]{Shao:2019gjj}%
  \BibitemOpen
  \bibfield  {author} {\bibinfo {author} {\bibfnamefont {L.}~\bibnamefont
  {Shao}},\ }\href {\doibase 10.1063/1.5117806} {\bibfield  {journal} {\bibinfo
   {journal} {AIP Conf. Proc.}\ }\textbf {\bibinfo {volume} {2127}},\ \bibinfo
  {pages} {020016} (\bibinfo {year} {2019}{\natexlab{b}})},\ \Eprint
  {http://arxiv.org/abs/1901.07546} {arXiv:1901.07546 [gr-qc]} \BibitemShut
  {NoStop}%
\bibitem [{\citenamefont {Tiglio}\ and\ \citenamefont
  {Villanueva}(2021)}]{Tiglio:2021ysj}%
  \BibitemOpen
  \bibfield  {author} {\bibinfo {author} {\bibfnamefont {M.}~\bibnamefont
  {Tiglio}}\ and\ \bibinfo {author} {\bibfnamefont {A.}~\bibnamefont
  {Villanueva}},\ }\href@noop {} {\  (\bibinfo {year} {2021})},\ \Eprint
  {http://arxiv.org/abs/2101.11608} {arXiv:2101.11608 [gr-qc]} \BibitemShut
  {NoStop}%
\bibitem [{\citenamefont {Field}\ \emph {et~al.}(2014)\citenamefont {Field},
  \citenamefont {Galley}, \citenamefont {Hesthaven}, \citenamefont {Kaye},\
  and\ \citenamefont {Tiglio}}]{Field2014PhRvX...4c1006F}%
  \BibitemOpen
  \bibfield  {author} {\bibinfo {author} {\bibfnamefont {S.~E.}\ \bibnamefont
  {Field}}, \bibinfo {author} {\bibfnamefont {C.~R.}\ \bibnamefont {Galley}},
  \bibinfo {author} {\bibfnamefont {J.~S.}\ \bibnamefont {Hesthaven}}, \bibinfo
  {author} {\bibfnamefont {J.}~\bibnamefont {Kaye}}, \ and\ \bibinfo {author}
  {\bibfnamefont {M.}~\bibnamefont {Tiglio}},\ }\href {\doibase
  10.1103/PhysRevX.4.031006} {\bibfield  {journal} {\bibinfo  {journal} {Phys.
  Rev. X}\ }\textbf {\bibinfo {volume} {4}},\ \bibinfo {pages} {031006}
  (\bibinfo {year} {2014})},\ \Eprint {http://arxiv.org/abs/1308.3565}
  {arXiv:1308.3565 [gr-qc]} \BibitemShut {NoStop}%
\bibitem [{\citenamefont {Canizares}\ \emph {et~al.}(2013)\citenamefont
  {Canizares}, \citenamefont {Field}, \citenamefont {Gair},\ and\ \citenamefont
  {Tiglio}}]{Canizares:2013ywa}%
  \BibitemOpen
  \bibfield  {author} {\bibinfo {author} {\bibfnamefont {P.}~\bibnamefont
  {Canizares}}, \bibinfo {author} {\bibfnamefont {S.~E.}\ \bibnamefont
  {Field}}, \bibinfo {author} {\bibfnamefont {J.~R.}\ \bibnamefont {Gair}}, \
  and\ \bibinfo {author} {\bibfnamefont {M.}~\bibnamefont {Tiglio}},\ }\href
  {\doibase 10.1103/PhysRevD.87.124005} {\bibfield  {journal} {\bibinfo
  {journal} {Phys. Rev. D}\ }\textbf {\bibinfo {volume} {87}},\ \bibinfo
  {pages} {124005} (\bibinfo {year} {2013})},\ \Eprint
  {http://arxiv.org/abs/1304.0462} {arXiv:1304.0462 [gr-qc]} \BibitemShut
  {NoStop}%
\bibitem [{\citenamefont {Canizares}\ \emph {et~al.}(2015)\citenamefont
  {Canizares}, \citenamefont {Field}, \citenamefont {Gair}, \citenamefont
  {Raymond}, \citenamefont {Smith},\ and\ \citenamefont
  {Tiglio}}]{Canizares:2014fya}%
  \BibitemOpen
  \bibfield  {author} {\bibinfo {author} {\bibfnamefont {P.}~\bibnamefont
  {Canizares}}, \bibinfo {author} {\bibfnamefont {S.~E.}\ \bibnamefont
  {Field}}, \bibinfo {author} {\bibfnamefont {J.}~\bibnamefont {Gair}},
  \bibinfo {author} {\bibfnamefont {V.}~\bibnamefont {Raymond}}, \bibinfo
  {author} {\bibfnamefont {R.}~\bibnamefont {Smith}}, \ and\ \bibinfo {author}
  {\bibfnamefont {M.}~\bibnamefont {Tiglio}},\ }\href {\doibase
  10.1103/PhysRevLett.114.071104} {\bibfield  {journal} {\bibinfo  {journal}
  {Phys. Rev. Lett.}\ }\textbf {\bibinfo {volume} {114}},\ \bibinfo {pages}
  {071104} (\bibinfo {year} {2015})},\ \Eprint {http://arxiv.org/abs/1404.6284}
  {arXiv:1404.6284 [gr-qc]} \BibitemShut {NoStop}%
\bibitem [{\citenamefont {Ramazano\u{g}lu}\ and\ \citenamefont
  {Pretorius}(2016)}]{Ramazanoglu:2016kul}%
  \BibitemOpen
  \bibfield  {author} {\bibinfo {author} {\bibfnamefont {F.~M.}\ \bibnamefont
  {Ramazano\u{g}lu}}\ and\ \bibinfo {author} {\bibfnamefont {F.}~\bibnamefont
  {Pretorius}},\ }\href {\doibase 10.1103/PhysRevD.93.064005} {\bibfield
  {journal} {\bibinfo  {journal} {Phys. Rev. D}\ }\textbf {\bibinfo {volume}
  {93}},\ \bibinfo {pages} {064005} (\bibinfo {year} {2016})},\ \Eprint
  {http://arxiv.org/abs/1601.07475} {arXiv:1601.07475 [gr-qc]} \BibitemShut
  {NoStop}%
\bibitem [{\citenamefont {Xu}\ \emph {et~al.}(2020)\citenamefont {Xu},
  \citenamefont {Gao},\ and\ \citenamefont {Shao}}]{Xu:2020vbs}%
  \BibitemOpen
  \bibfield  {author} {\bibinfo {author} {\bibfnamefont {R.}~\bibnamefont
  {Xu}}, \bibinfo {author} {\bibfnamefont {Y.}~\bibnamefont {Gao}}, \ and\
  \bibinfo {author} {\bibfnamefont {L.}~\bibnamefont {Shao}},\ }\href {\doibase
  10.1103/PhysRevD.102.064057} {\bibfield  {journal} {\bibinfo  {journal}
  {Phys. Rev. D}\ }\textbf {\bibinfo {volume} {102}},\ \bibinfo {pages}
  {064057} (\bibinfo {year} {2020})},\ \Eprint
  {http://arxiv.org/abs/2007.10080} {arXiv:2007.10080 [gr-qc]} \BibitemShut
  {NoStop}%
\bibitem [{\citenamefont {Barausse}\ \emph {et~al.}(2013)\citenamefont
  {Barausse}, \citenamefont {Palenzuela}, \citenamefont {Ponce},\ and\
  \citenamefont {Lehner}}]{Barausse2013PhysRevD.87.081506}%
  \BibitemOpen
  \bibfield  {author} {\bibinfo {author} {\bibfnamefont {E.}~\bibnamefont
  {Barausse}}, \bibinfo {author} {\bibfnamefont {C.}~\bibnamefont
  {Palenzuela}}, \bibinfo {author} {\bibfnamefont {M.}~\bibnamefont {Ponce}}, \
  and\ \bibinfo {author} {\bibfnamefont {L.}~\bibnamefont {Lehner}},\ }\href
  {\doibase 10.1103/PhysRevD.87.081506} {\bibfield  {journal} {\bibinfo
  {journal} {Phys. Rev. D}\ }\textbf {\bibinfo {volume} {87}},\ \bibinfo
  {pages} {081506} (\bibinfo {year} {2013})}\BibitemShut {NoStop}%
\bibitem [{\citenamefont {Kramer}\ \emph {et~al.}(2006)\citenamefont {Kramer}
  \emph {et~al.}}]{Kramer2006Sci...314...97K}%
  \BibitemOpen
  \bibfield  {author} {\bibinfo {author} {\bibfnamefont {M.}~\bibnamefont
  {Kramer}} \emph {et~al.},\ }\href {\doibase 10.1126/science.1132305}
  {\bibfield  {journal} {\bibinfo  {journal} {Science}\ }\textbf {\bibinfo
  {volume} {314}},\ \bibinfo {pages} {97} (\bibinfo {year} {2006})},\ \Eprint
  {http://arxiv.org/abs/astro-ph/0609417} {arXiv:astro-ph/0609417} \BibitemShut
  {NoStop}%
\bibitem [{\citenamefont {Fonseca}\ \emph {et~al.}(2021)\citenamefont {Fonseca}
  \emph {et~al.}}]{Fonseca:2021wxt}%
  \BibitemOpen
  \bibfield  {author} {\bibinfo {author} {\bibfnamefont {E.}~\bibnamefont
  {Fonseca}} \emph {et~al.},\ }\href {\doibase 10.3847/2041-8213/ac03b8}
  {\bibfield  {journal} {\bibinfo  {journal} {Astrophys. J. Lett.}\ }\textbf
  {\bibinfo {volume} {915}},\ \bibinfo {pages} {L12} (\bibinfo {year}
  {2021})},\ \Eprint {http://arxiv.org/abs/2104.00880} {arXiv:2104.00880
  [astro-ph.HE]} \BibitemShut {NoStop}%
\bibitem [{\citenamefont {Barrault}\ \emph {et~al.}(2004)\citenamefont
  {Barrault}, \citenamefont {Maday}, \citenamefont {Nguyen},\ and\
  \citenamefont {Patera}}]{BARRAULT2004667}%
  \BibitemOpen
  \bibfield  {author} {\bibinfo {author} {\bibfnamefont {M.}~\bibnamefont
  {Barrault}}, \bibinfo {author} {\bibfnamefont {Y.}~\bibnamefont {Maday}},
  \bibinfo {author} {\bibfnamefont {N.~C.}\ \bibnamefont {Nguyen}}, \ and\
  \bibinfo {author} {\bibfnamefont {A.~T.}\ \bibnamefont {Patera}},\ }\href
  {\doibase https://doi.org/10.1016/j.crma.2004.08.006} {\bibfield  {journal}
  {\bibinfo  {journal} {Comptes Rendus Mathematique}\ }\textbf {\bibinfo
  {volume} {339}},\ \bibinfo {pages} {667} (\bibinfo {year}
  {2004})}\BibitemShut {NoStop}%
\bibitem [{\citenamefont {Antoniadis}\ \emph {et~al.}(2013)\citenamefont
  {Antoniadis} \emph {et~al.}}]{Antoniadis:2013pzd}%
  \BibitemOpen
  \bibfield  {author} {\bibinfo {author} {\bibfnamefont {J.}~\bibnamefont
  {Antoniadis}} \emph {et~al.},\ }\href {\doibase 10.1126/science.1233232}
  {\bibfield  {journal} {\bibinfo  {journal} {Science}\ }\textbf {\bibinfo
  {volume} {340}},\ \bibinfo {pages} {6131} (\bibinfo {year} {2013})},\ \Eprint
  {http://arxiv.org/abs/1304.6875} {arXiv:1304.6875 [astro-ph.HE]} \BibitemShut
  {NoStop}%
\bibitem [{\citenamefont {Lazaridis}\ \emph {et~al.}(2009)\citenamefont
  {Lazaridis} \emph {et~al.}}]{Lazaridis2009MNRAS.400..805L}%
  \BibitemOpen
  \bibfield  {author} {\bibinfo {author} {\bibfnamefont {K.}~\bibnamefont
  {Lazaridis}} \emph {et~al.},\ }\href {\doibase
  10.1111/j.1365-2966.2009.15481.x} {\bibfield  {journal} {\bibinfo  {journal}
  {Mon. Not. R. Astron. Soc.}\ }\textbf {\bibinfo {volume} {400}},\ \bibinfo
  {pages} {805} (\bibinfo {year} {2009})},\ \Eprint
  {http://arxiv.org/abs/0908.0285} {arXiv:0908.0285 [astro-ph.GA]} \BibitemShut
  {NoStop}%
\bibitem [{\citenamefont {Desvignes}\ \emph {et~al.}(2016)\citenamefont
  {Desvignes} \emph {et~al.}}]{Desvignes2016MNRAS.458.3341D}%
  \BibitemOpen
  \bibfield  {author} {\bibinfo {author} {\bibfnamefont {G.}~\bibnamefont
  {Desvignes}} \emph {et~al.},\ }\href {\doibase 10.1093/mnras/stw483}
  {\bibfield  {journal} {\bibinfo  {journal} {Mon. Not. Roy. Astron. Soc.}\
  }\textbf {\bibinfo {volume} {458}},\ \bibinfo {pages} {3341} (\bibinfo {year}
  {2016})},\ \Eprint {http://arxiv.org/abs/1602.08511} {arXiv:1602.08511
  [astro-ph.HE]} \BibitemShut {NoStop}%
\bibitem [{\citenamefont {Antoniadis}\ \emph {et~al.}(2016)\citenamefont
  {Antoniadis}, \citenamefont {Tauris}, \citenamefont {{\"O}zel}, \citenamefont
  {Barr}, \citenamefont {Champion},\ and\ \citenamefont
  {Freire}}]{Antoniadis2016arXiv160501665A}%
  \BibitemOpen
  \bibfield  {author} {\bibinfo {author} {\bibfnamefont {J.}~\bibnamefont
  {Antoniadis}}, \bibinfo {author} {\bibfnamefont {T.~M.}\ \bibnamefont
  {Tauris}}, \bibinfo {author} {\bibfnamefont {F.}~\bibnamefont {{\"O}zel}},
  \bibinfo {author} {\bibfnamefont {E.}~\bibnamefont {Barr}}, \bibinfo {author}
  {\bibfnamefont {D.~J.}\ \bibnamefont {Champion}}, \ and\ \bibinfo {author}
  {\bibfnamefont {P.~C.~C.}\ \bibnamefont {Freire}},\ }\href@noop {} {\
  (\bibinfo {year} {2016})},\ \Eprint {http://arxiv.org/abs/1605.01665}
  {arXiv:1605.01665 [astro-ph.HE]} \BibitemShut {NoStop}%
\bibitem [{\citenamefont {{Mata S{\'a}nchez}}\ \emph
  {et~al.}(2020)\citenamefont {{Mata S{\'a}nchez}}, \citenamefont {{Istrate}},
  \citenamefont {{van Kerkwijk}}, \citenamefont {{Breton}},\ and\ \citenamefont
  {{Kaplan}}}]{Sanchez2020MNRAS.494.4031M}%
  \BibitemOpen
  \bibfield  {author} {\bibinfo {author} {\bibfnamefont {D.}~\bibnamefont
  {{Mata S{\'a}nchez}}}, \bibinfo {author} {\bibfnamefont {A.~G.}\ \bibnamefont
  {{Istrate}}}, \bibinfo {author} {\bibfnamefont {M.~H.}\ \bibnamefont {{van
  Kerkwijk}}}, \bibinfo {author} {\bibfnamefont {R.~P.}\ \bibnamefont
  {{Breton}}}, \ and\ \bibinfo {author} {\bibfnamefont {D.~L.}\ \bibnamefont
  {{Kaplan}}},\ }\href {\doibase 10.1093/mnras/staa983} {\bibfield  {journal}
  {\bibinfo  {journal} {\mnras}\ }\textbf {\bibinfo {volume} {494}},\ \bibinfo
  {pages} {4031} (\bibinfo {year} {2020})},\ \Eprint
  {http://arxiv.org/abs/2004.02901} {arXiv:2004.02901 [astro-ph.HE]}
  \BibitemShut {NoStop}%
\bibitem [{\citenamefont {Freire}\ \emph {et~al.}(2012)\citenamefont {Freire},
  \citenamefont {Wex}, \citenamefont {Esposito-Far\`ese}, \citenamefont
  {Verbiest}, \citenamefont {Bailes}, \citenamefont {Jacoby}, \citenamefont
  {Kramer}, \citenamefont {Stairs}, \citenamefont {Antoniadis},\ and\
  \citenamefont {Janssen}}]{Freire2012MNRAS.423.3328F}%
  \BibitemOpen
  \bibfield  {author} {\bibinfo {author} {\bibfnamefont {P.~C.~C.}\
  \bibnamefont {Freire}}, \bibinfo {author} {\bibfnamefont {N.}~\bibnamefont
  {Wex}}, \bibinfo {author} {\bibfnamefont {G.}~\bibnamefont
  {Esposito-Far\`ese}}, \bibinfo {author} {\bibfnamefont {J.~P.~W.}\
  \bibnamefont {Verbiest}}, \bibinfo {author} {\bibfnamefont {M.}~\bibnamefont
  {Bailes}}, \bibinfo {author} {\bibfnamefont {B.~A.}\ \bibnamefont {Jacoby}},
  \bibinfo {author} {\bibfnamefont {M.}~\bibnamefont {Kramer}}, \bibinfo
  {author} {\bibfnamefont {I.~H.}\ \bibnamefont {Stairs}}, \bibinfo {author}
  {\bibfnamefont {J.}~\bibnamefont {Antoniadis}}, \ and\ \bibinfo {author}
  {\bibfnamefont {G.~H.}\ \bibnamefont {Janssen}},\ }\href {\doibase
  10.1111/j.1365-2966.2012.21253.x} {\bibfield  {journal} {\bibinfo  {journal}
  {Mon. Not. Roy. Astron. Soc.}\ }\textbf {\bibinfo {volume} {423}},\ \bibinfo
  {pages} {3328} (\bibinfo {year} {2012})},\ \Eprint
  {http://arxiv.org/abs/1205.1450} {arXiv:1205.1450 [astro-ph.GA]} \BibitemShut
  {NoStop}%
\bibitem [{\citenamefont {Liu}\ \emph {et~al.}(2020{\natexlab{b}})\citenamefont
  {Liu} \emph {et~al.}}]{Liu2020MNRAS.499.2276L}%
  \BibitemOpen
  \bibfield  {author} {\bibinfo {author} {\bibfnamefont {K.}~\bibnamefont
  {Liu}} \emph {et~al.},\ }\href {\doibase 10.1093/mnras/staa2993} {\bibfield
  {journal} {\bibinfo  {journal} {Mon. Not. Roy. Astron. Soc.}\ }\textbf
  {\bibinfo {volume} {499}},\ \bibinfo {pages} {2276} (\bibinfo {year}
  {2020}{\natexlab{b}})},\ \Eprint {http://arxiv.org/abs/2009.12544}
  {arXiv:2009.12544 [astro-ph.HE]} \BibitemShut {NoStop}%
\bibitem [{\citenamefont {Cognard}\ \emph {et~al.}(2017)\citenamefont {Cognard}
  \emph {et~al.}}]{Cognard2017ApJ...844..128C}%
  \BibitemOpen
  \bibfield  {author} {\bibinfo {author} {\bibfnamefont {I.}~\bibnamefont
  {Cognard}} \emph {et~al.},\ }\href {\doibase 10.3847/1538-4357/aa7bee}
  {\bibfield  {journal} {\bibinfo  {journal} {Astrophys. J.}\ }\textbf
  {\bibinfo {volume} {844}},\ \bibinfo {pages} {128} (\bibinfo {year}
  {2017})},\ \Eprint {http://arxiv.org/abs/1706.08060} {arXiv:1706.08060
  [astro-ph.HE]} \BibitemShut {NoStop}%
\bibitem [{\citenamefont {Damour}\ and\ \citenamefont
  {Taylor}(1991)}]{Damour:1990wz}%
  \BibitemOpen
  \bibfield  {author} {\bibinfo {author} {\bibfnamefont {T.}~\bibnamefont
  {Damour}}\ and\ \bibinfo {author} {\bibfnamefont {J.~H.}\ \bibnamefont
  {Taylor}},\ }\href {\doibase 10.1086/169585} {\bibfield  {journal} {\bibinfo
  {journal} {Astrophys. J.}\ }\textbf {\bibinfo {volume} {366}},\ \bibinfo
  {pages} {501} (\bibinfo {year} {1991})}\BibitemShut {NoStop}%
\bibitem [{\citenamefont {{Shklovskii}}(1970)}]{Shklovskii1970SvA....13..562S}%
  \BibitemOpen
  \bibfield  {author} {\bibinfo {author} {\bibfnamefont {I.~S.}\ \bibnamefont
  {{Shklovskii}}},\ }\href
  {https://ui.adsabs.harvard.edu/abs/1970SvA....13..562S} {\bibfield  {journal}
  {\bibinfo  {journal} {Soviet Astronomy}\ }\textbf {\bibinfo {volume} {13}},\
  \bibinfo {pages} {562} (\bibinfo {year} {1970})}\BibitemShut {NoStop}%
\bibitem [{\citenamefont {{McMillan}}(2017)}]{McMillan2017MNRAS.465...76M}%
  \BibitemOpen
  \bibfield  {author} {\bibinfo {author} {\bibfnamefont {P.~J.}\ \bibnamefont
  {{McMillan}}},\ }\href {\doibase 10.1093/mnras/stw2759} {\bibfield  {journal}
  {\bibinfo  {journal} {Mon. Not. Roy. Astron. Soc.}\ }\textbf {\bibinfo
  {volume} {465}},\ \bibinfo {pages} {76} (\bibinfo {year} {2017})},\ \Eprint
  {http://arxiv.org/abs/1608.00971} {arXiv:1608.00971 [astro-ph.GA]}
  \BibitemShut {NoStop}%
\bibitem [{\citenamefont {{Weisberg}}\ and\ \citenamefont
  {{Huang}}(2016)}]{Weisberg2016ApJ...829...55W}%
  \BibitemOpen
  \bibfield  {author} {\bibinfo {author} {\bibfnamefont {J.~M.}\ \bibnamefont
  {{Weisberg}}}\ and\ \bibinfo {author} {\bibfnamefont {Y.}~\bibnamefont
  {{Huang}}},\ }\href {\doibase 10.3847/0004-637X/829/1/55} {\bibfield
  {journal} {\bibinfo  {journal} {\apj}\ }\textbf {\bibinfo {volume} {829}},\
  \bibinfo {eid} {55} (\bibinfo {year} {2016})},\ \Eprint
  {http://arxiv.org/abs/1606.02744} {arXiv:1606.02744 [astro-ph.HE]}
  \BibitemShut {NoStop}%
\bibitem [{\citenamefont {Cameron}\ \emph {et~al.}(2018)\citenamefont {Cameron}
  \emph {et~al.}}]{Cameron2018MNRAS.475L..57C}%
  \BibitemOpen
  \bibfield  {author} {\bibinfo {author} {\bibfnamefont {A.~D.}\ \bibnamefont
  {Cameron}} \emph {et~al.},\ }\href {\doibase 10.1093/mnrasl/sly003}
  {\bibfield  {journal} {\bibinfo  {journal} {Mon. Not. Roy. Astron. Soc.}\
  }\textbf {\bibinfo {volume} {475}},\ \bibinfo {pages} {L57} (\bibinfo {year}
  {2018})},\ \Eprint {http://arxiv.org/abs/1711.07697} {arXiv:1711.07697
  [astro-ph.HE]} \BibitemShut {NoStop}%
\bibitem [{\citenamefont {Foreman-Mackey}\ \emph {et~al.}(2013)\citenamefont
  {Foreman-Mackey}, \citenamefont {Hogg}, \citenamefont {Lang},\ and\
  \citenamefont {Goodman}}]{ForemanMackey:2012ig}%
  \BibitemOpen
  \bibfield  {author} {\bibinfo {author} {\bibfnamefont {D.}~\bibnamefont
  {Foreman-Mackey}}, \bibinfo {author} {\bibfnamefont {D.~W.}\ \bibnamefont
  {Hogg}}, \bibinfo {author} {\bibfnamefont {D.}~\bibnamefont {Lang}}, \ and\
  \bibinfo {author} {\bibfnamefont {J.}~\bibnamefont {Goodman}},\ }\href
  {\doibase 10.1086/670067} {\bibfield  {journal} {\bibinfo  {journal} {Publ.
  Astron. Soc. Pac.}\ }\textbf {\bibinfo {volume} {125}},\ \bibinfo {pages}
  {306} (\bibinfo {year} {2013})},\ \Eprint {http://arxiv.org/abs/1202.3665}
  {arXiv:1202.3665 [astro-ph.IM]} \BibitemShut {NoStop}%
\bibitem [{\citenamefont {{Taylor}}(1987)}]{1987grg..conf..209T}%
  \BibitemOpen
  \bibfield  {author} {\bibinfo {author} {\bibfnamefont {J.~H.}\ \bibnamefont
  {{Taylor}}},\ }in\ \href@noop {} {\emph {\bibinfo {booktitle} {General
  Relativity and Gravitation}}}\ (\bibinfo {year} {1987})\ pp.\ \bibinfo
  {pages} {209--222}\BibitemShut {NoStop}%
\bibitem [{\citenamefont {Taylor}\ and\ \citenamefont
  {Weisberg}(1989)}]{Taylor:1989sw}%
  \BibitemOpen
  \bibfield  {author} {\bibinfo {author} {\bibfnamefont {J.~H.}\ \bibnamefont
  {Taylor}}\ and\ \bibinfo {author} {\bibfnamefont {J.~M.}\ \bibnamefont
  {Weisberg}},\ }\href {\doibase 10.1086/167917} {\bibfield  {journal}
  {\bibinfo  {journal} {Astrophys. J.}\ }\textbf {\bibinfo {volume} {345}},\
  \bibinfo {pages} {434} (\bibinfo {year} {1989})}\BibitemShut {NoStop}%
\bibitem [{\citenamefont {{Lattimer}}(2012)}]{Lattimer2012ARNPS..62..485L}%
  \BibitemOpen
  \bibfield  {author} {\bibinfo {author} {\bibfnamefont {J.~M.}\ \bibnamefont
  {{Lattimer}}},\ }\href {\doibase 10.1146/annurev-nucl-102711-095018}
  {\bibfield  {journal} {\bibinfo  {journal} {Annual Review of Nuclear and
  Particle Science}\ }\textbf {\bibinfo {volume} {62}},\ \bibinfo {pages} {485}
  (\bibinfo {year} {2012})},\ \Eprint {http://arxiv.org/abs/1305.3510}
  {arXiv:1305.3510 [nucl-th]} \BibitemShut {NoStop}%
\bibitem [{\citenamefont {De}\ \emph {et~al.}(2018)\citenamefont {De},
  \citenamefont {Finstad}, \citenamefont {Lattimer}, \citenamefont {Brown},
  \citenamefont {Berger},\ and\ \citenamefont {Biwer}}]{De:2018uhw}%
  \BibitemOpen
  \bibfield  {author} {\bibinfo {author} {\bibfnamefont {S.}~\bibnamefont
  {De}}, \bibinfo {author} {\bibfnamefont {D.}~\bibnamefont {Finstad}},
  \bibinfo {author} {\bibfnamefont {J.~M.}\ \bibnamefont {Lattimer}}, \bibinfo
  {author} {\bibfnamefont {D.~A.}\ \bibnamefont {Brown}}, \bibinfo {author}
  {\bibfnamefont {E.}~\bibnamefont {Berger}}, \ and\ \bibinfo {author}
  {\bibfnamefont {C.~M.}\ \bibnamefont {Biwer}},\ }\href {\doibase
  10.1103/PhysRevLett.121.091102} {\bibfield  {journal} {\bibinfo  {journal}
  {Phys. Rev. Lett.}\ }\textbf {\bibinfo {volume} {121}},\ \bibinfo {pages}
  {091102} (\bibinfo {year} {2018})},\ \bibinfo {note} {[Erratum:
  Phys.Rev.Lett. 121, 259902 (2018)]},\ \Eprint
  {http://arxiv.org/abs/1804.08583} {arXiv:1804.08583 [astro-ph.HE]}
  \BibitemShut {NoStop}%
\bibitem [{\citenamefont {Capano}\ \emph {et~al.}(2020)\citenamefont {Capano},
  \citenamefont {Tews}, \citenamefont {Brown}, \citenamefont {Margalit},
  \citenamefont {De}, \citenamefont {Kumar}, \citenamefont {Brown},
  \citenamefont {Krishnan},\ and\ \citenamefont {Reddy}}]{Capano:2019eae}%
  \BibitemOpen
  \bibfield  {author} {\bibinfo {author} {\bibfnamefont {C.~D.}\ \bibnamefont
  {Capano}}, \bibinfo {author} {\bibfnamefont {I.}~\bibnamefont {Tews}},
  \bibinfo {author} {\bibfnamefont {S.~M.}\ \bibnamefont {Brown}}, \bibinfo
  {author} {\bibfnamefont {B.}~\bibnamefont {Margalit}}, \bibinfo {author}
  {\bibfnamefont {S.}~\bibnamefont {De}}, \bibinfo {author} {\bibfnamefont
  {S.}~\bibnamefont {Kumar}}, \bibinfo {author} {\bibfnamefont {D.~A.}\
  \bibnamefont {Brown}}, \bibinfo {author} {\bibfnamefont {B.}~\bibnamefont
  {Krishnan}}, \ and\ \bibinfo {author} {\bibfnamefont {S.}~\bibnamefont
  {Reddy}},\ }\href {\doibase 10.1038/s41550-020-1014-6} {\bibfield  {journal}
  {\bibinfo  {journal} {Nature Astron.}\ }\textbf {\bibinfo {volume} {4}},\
  \bibinfo {pages} {625} (\bibinfo {year} {2020})},\ \Eprint
  {http://arxiv.org/abs/1908.10352} {arXiv:1908.10352 [astro-ph.HE]}
  \BibitemShut {NoStop}%
\bibitem [{\citenamefont {Del~Pozzo}\ and\ \citenamefont
  {Vecchio}(2016)}]{DelPozzo:2016ugt}%
  \BibitemOpen
  \bibfield  {author} {\bibinfo {author} {\bibfnamefont {W.}~\bibnamefont
  {Del~Pozzo}}\ and\ \bibinfo {author} {\bibfnamefont {A.}~\bibnamefont
  {Vecchio}},\ }\href {\doibase 10.1093/mnrasl/slw116} {\bibfield  {journal}
  {\bibinfo  {journal} {Mon. Not. Roy. Astron. Soc.}\ }\textbf {\bibinfo
  {volume} {462}},\ \bibinfo {pages} {L21} (\bibinfo {year} {2016})},\ \Eprint
  {http://arxiv.org/abs/1606.02852} {arXiv:1606.02852 [gr-qc]} \BibitemShut
  {NoStop}%
\bibitem [{\citenamefont {Gelman}\ and\ \citenamefont
  {Rubin}(1992)}]{Gelman:1992zz}%
  \BibitemOpen
  \bibfield  {author} {\bibinfo {author} {\bibfnamefont {A.}~\bibnamefont
  {Gelman}}\ and\ \bibinfo {author} {\bibfnamefont {D.~B.}\ \bibnamefont
  {Rubin}},\ }\href {\doibase 10.1214/ss/1177011136} {\bibfield  {journal}
  {\bibinfo  {journal} {Statist. Sci.}\ }\textbf {\bibinfo {volume} {7}},\
  \bibinfo {pages} {457} (\bibinfo {year} {1992})}\BibitemShut {NoStop}%
\bibitem [{\citenamefont {Nan}\ \emph {et~al.}(2011)\citenamefont {Nan},
  \citenamefont {Li}, \citenamefont {Jin}, \citenamefont {Wang}, \citenamefont
  {Zhu}, \citenamefont {Zhu}, \citenamefont {Zhang}, \citenamefont {Yue},\ and\
  \citenamefont {Qian}}]{Nan:2011um}%
  \BibitemOpen
  \bibfield  {author} {\bibinfo {author} {\bibfnamefont {R.}~\bibnamefont
  {Nan}}, \bibinfo {author} {\bibfnamefont {D.}~\bibnamefont {Li}}, \bibinfo
  {author} {\bibfnamefont {C.}~\bibnamefont {Jin}}, \bibinfo {author}
  {\bibfnamefont {Q.}~\bibnamefont {Wang}}, \bibinfo {author} {\bibfnamefont
  {L.}~\bibnamefont {Zhu}}, \bibinfo {author} {\bibfnamefont {W.}~\bibnamefont
  {Zhu}}, \bibinfo {author} {\bibfnamefont {H.}~\bibnamefont {Zhang}}, \bibinfo
  {author} {\bibfnamefont {Y.}~\bibnamefont {Yue}}, \ and\ \bibinfo {author}
  {\bibfnamefont {L.}~\bibnamefont {Qian}},\ }\href {\doibase
  10.1142/S0218271811019335} {\bibfield  {journal} {\bibinfo  {journal} {Int.
  J. Mod. Phys. D}\ }\textbf {\bibinfo {volume} {20}},\ \bibinfo {pages} {989}
  (\bibinfo {year} {2011})},\ \Eprint {http://arxiv.org/abs/1105.3794}
  {arXiv:1105.3794 [astro-ph.IM]} \BibitemShut {NoStop}%
\bibitem [{\citenamefont {Kramer}\ \emph {et~al.}(2004)\citenamefont {Kramer},
  \citenamefont {Backer}, \citenamefont {Cordes}, \citenamefont {Lazio},
  \citenamefont {Stappers},\ and\ \citenamefont {Johnston}}]{Kramer:2004hd}%
  \BibitemOpen
  \bibfield  {author} {\bibinfo {author} {\bibfnamefont {M.}~\bibnamefont
  {Kramer}}, \bibinfo {author} {\bibfnamefont {D.~C.}\ \bibnamefont {Backer}},
  \bibinfo {author} {\bibfnamefont {J.~M.}\ \bibnamefont {Cordes}}, \bibinfo
  {author} {\bibfnamefont {T.~J.~W.}\ \bibnamefont {Lazio}}, \bibinfo {author}
  {\bibfnamefont {B.~W.}\ \bibnamefont {Stappers}}, \ and\ \bibinfo {author}
  {\bibfnamefont {S.}~\bibnamefont {Johnston}},\ }\href {\doibase
  10.1016/j.newar.2004.09.020} {\bibfield  {journal} {\bibinfo  {journal} {New
  Astron. Rev.}\ }\textbf {\bibinfo {volume} {48}},\ \bibinfo {pages} {993}
  (\bibinfo {year} {2004})},\ \Eprint {http://arxiv.org/abs/astro-ph/0409379}
  {arXiv:astro-ph/0409379} \BibitemShut {NoStop}%
\bibitem [{\citenamefont {Shao}\ \emph {et~al.}(2015)\citenamefont {Shao} \emph
  {et~al.}}]{Shao:2014wja}%
  \BibitemOpen
  \bibfield  {author} {\bibinfo {author} {\bibfnamefont {L.}~\bibnamefont
  {Shao}} \emph {et~al.},\ }\href {\doibase 10.22323/1.215.0042} {\bibfield
  {journal} {\bibinfo  {journal} {PoS}\ }\textbf {\bibinfo {volume}
  {AASKA14}},\ \bibinfo {pages} {042} (\bibinfo {year} {2015})},\ \Eprint
  {http://arxiv.org/abs/1501.00058} {arXiv:1501.00058 [astro-ph.HE]}
  \BibitemShut {NoStop}%
\end{thebibliography}%

\end{document}